\documentclass[aps,twocolumn,preprintnumbers,amsmath,amssymb,nofootinbib,superscriptaddress,notitlepage]{revtex4}
\usepackage[sort&compress]{natbib}
\usepackage{graphicx,color,dcolumn,booktabs,bm}
\usepackage{longtable,lscape}
\usepackage{txfonts}
\usepackage{overpic}
\usepackage{amssymb}
\usepackage{indentfirst}
\usepackage{feynmf}   
\usepackage{slashed}  
\usepackage{cases}
\usepackage{color}
\usepackage{multirow}
\usepackage{longtable}
\usepackage{makecell}
\usepackage{ulem}
\usepackage{enumerate}
\usepackage{graphicx,color,dcolumn,booktabs,bm}
\usepackage[colorlinks,
            citecolor=blue,
            anchorcolor=red,
            menucolor=red,
            linkcolor=red,
            filecolor=red,
            runcolor=red,
            urlcolor=blue,
            frenchlinks=red]{hyperref}
\usepackage{epstopdf}

\begin{document}

\title{Triple-charm molecular states composed of $D^*D^*D$ and $D^*D^*D^*$}

\author{Si-Qiang Luo}
\affiliation{School of Physical Science and Technology, Lanzhou University, Lanzhou 730000, China}
\affiliation{School of mathematics and statistics, Lanzhou University, Lanzhou 730000, China}

\author{Tian-Wei Wu}
\affiliation{School of Fundamental Physics and Mathematical Sciences, Hangzhou Institute for Advanced Study, UCAS, Hangzhou, 310024, China}

\author{Ming-Zhu Liu}
\affiliation{School of Space and Environment, Beihang University, Beijing 102206, China}
\affiliation{School of Physics, Beihang University, Beijing 102206, China}

\author{Li-Sheng Geng}\email{lisheng.geng@buaa.edu.cn}
\affiliation{School of Physics, Beihang University, Beijing 102206, China}
\affiliation{Beijing Key Laboratory of Advanced Nuclear Materials and Physics, Beihang University, Beijing, 102206, China}
\affiliation{School of Physics and Microelectronics, Zhengzhou University, Zhengzhou, Henan 450001, China}
\affiliation{Lanzhou Center for Theoretical Physics, Lanzhou University, Lanzhou 730000, China}

\author{Xiang Liu}\email{xiangliu@lzu.edu.cn}
\affiliation{School of Physical Science and Technology, Lanzhou University, Lanzhou 730000, China}
\affiliation{Lanzhou Center for Theoretical Physics, Lanzhou University, Lanzhou 730000, China}
\affiliation{Research Center for Hadron and CSR Physics, Lanzhou University and Institute of Modern Physics of CAS, Lanzhou 730000, China}
\affiliation{Key Laboratory of Theoretical Physics of Gansu Province, and Frontiers Science Center for Rare Isotopes, Lanzhou University, Lanzhou 730000, China}

\begin{abstract}
Inspired by the newly observed $T_{cc}^+$ state, we systematically investigate the $S$-wave triple-charm molecular states composed of $D^*D^*D$ and $D^*D^*D^*$. We employ the one-boson-exchange model to derive the interactions between $D(D^*)$ and $D^*$ and solve the three-body Schr\"odinger equations with the Gaussian expansion method. The $S$-$D$ mixing and coupled channel effects are carefully assessed in our study. Our results show that the  $I(J^P)=\frac{1}{2}(0^-,1^-,2^-)$ $D^*D^*D$ and $I(J^P)=\frac{1}{2}(0^-,1^-,2^-,3^-)$ $D^*D^*D^*$ systems could form bound states, which can be viewed as three-body hadronic molecules. We  present not only the binding energies of the three-body bound states, but also the root-mean-square radii of $D $-$D^*$ and $D^*$-$D^*$, which further corroborate the molecular nature of these states. These predictions could be tested in the future at LHC or HL-LHC.
\end{abstract}

\maketitle

\section{Introduction}

As important members of the hadron family, exotic states have always interested both  theoreticians  and experimentalists. By definition, exotic states contain more complex quark and gluon contents than the conventional $q\bar{q}$ mesons and $qqq$ baryons. Given their peculiar nature, studies of exotic states have been  a hot topic in hadron physics.

Among the various exotic states, hadronic molecules  are quite distinct. They are loosely bound states composed of two or several conventional hadrons and provide a good laboratory to study hadron structure and nonperturbative strong interactions at hadronic level. In 2003, the $BABAR$ Collaboration observed a charmed-strange state $D_{s0}^*(2317)$ in the $D_s\pi^0$ channel~\cite{BaBar:2003oey}. Soon after, the CLEO Collaboration not only confirmed its existence , but also found a new charmed-strange state $D_{s1}^\prime(2460)$ in the $D_s^*\pi^0$ mass spectrum~\cite{CLEO:2003ggt}. In the same year, the Belle Collaboration reported a hidden-charm state $X(3872)$ in the $J/\psi\pi^+\pi^-$ channel~\cite{Belle:2003nnu}. The $D_{s0}^*(2317)$, $D_{s1}^\prime(2460)$, and $X(3872)$ states have two peculiar features. The first is that their masses are about 100 MeV below the potential model predictions, which implies that it is difficult to categorize them as conventional mesons. The second  is that $D_{s0}^*(2317)$, $D_{s1}^\prime(2460)$, and $X(3872)$ are close to and lower than the $DK$, $D^*K$, and $D\bar{D}^*$ thresholds, which strongly hints at their molecular nature. Although there are still many controversies, hadronic molecules are one of the most popular interpretations of these exotic hadrons~\cite{Barnes:2003dj,Xie:2010zza,Guo:2006fu,Faessler:2007gv,Feng:2012zze,Faessler:2007us,Zhang:2006ix,Swanson:2003tb,Wong:2003xk,Liu:2009qhy,Lee:2009hy,Altenbuchinger:2013vwa}. The observations of $D_{s0}^*(2317)$, $D_{s1}^\prime(2460)$, and $X(3872)$ opened a new era in searches for exotic states. In the following years, a plethora of hidden-charm $XYZ$ and $P_c$ states were observed in experiments (for reviews, see Refs.~\cite{Chen:2016qju,Liu:2013waa,Yuan:2018inv,Olsen:2017bmm,Guo:2017jvc,Hosaka:2016pey,Brambilla:2019esw}).

Very recently, the LHCb Collaboration observed a $T_{cc}^+$ state in the $D^0D^0\pi^+$ channel~\cite{LHCb:2021vvq,LHCb:2021auc}, whose mass and width obtained from a Breit-Wigner fit are
\begin{equation}
\begin{split}
m_{\rm BW}=&(m_{D^{*+}}+m_{D^0})-273\pm61\pm5^{+11}_{-14}\;{\rm keV},\\
\Gamma_{\rm BW}=&410\pm165\pm43^{+18}_{-38}\;{\rm keV}.
\end{split}
\end{equation}
On the other hand, the pole position is given as~\cite{LHCb:2021auc}
\begin{equation}
\begin{split}
m_{\rm pole}=&(m_{D^{*+}}+m_{D^0})-360\pm40^{+4}_{-0}\;{\rm keV},\\
\Gamma_{\rm pole}=&48\pm2^{+0}_{-14}\;{\rm keV}.
\end{split}
\end{equation}

\begin{figure}[htbp]
\includegraphics[width=8.6cm]{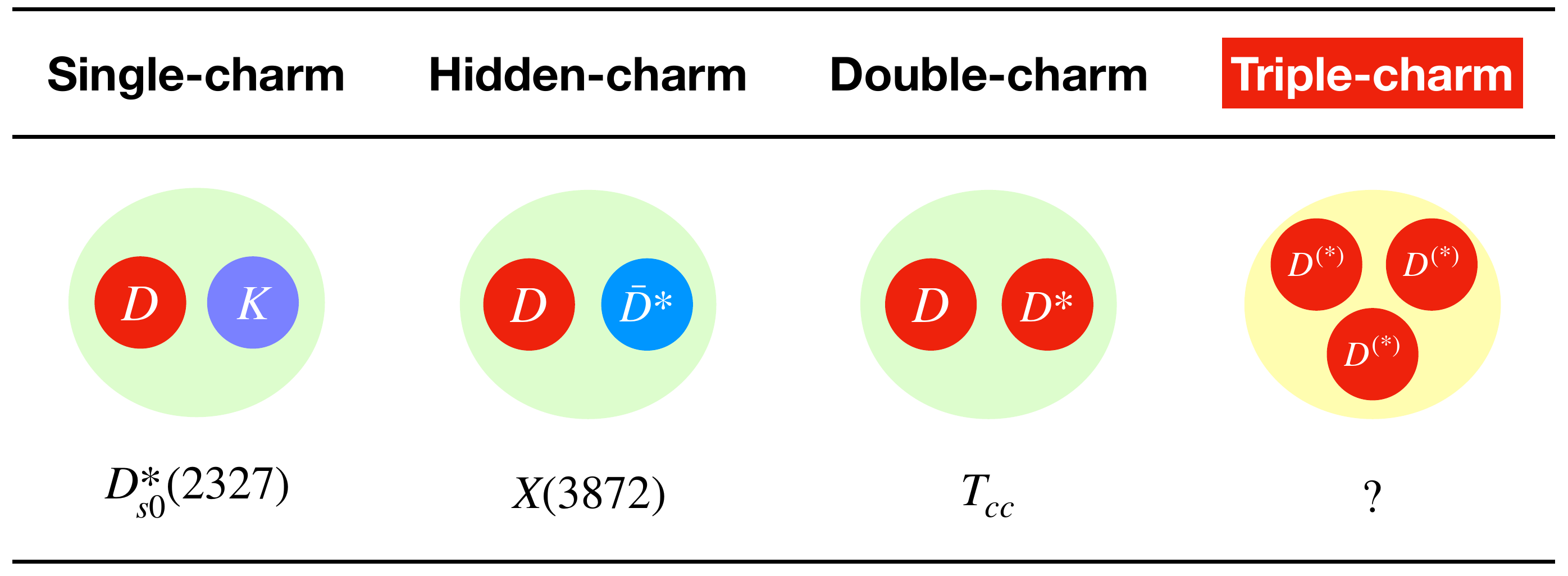}
\caption{Various types of hadronic molecular candidates. Here, we choose  $D_{s0}^*(2317)$, $X(3872)$, and $T_{cc}$ as examples.}
\label{moleculartype}
\end{figure}

From the decay products of $T_{cc}^+$, one can infer its minimum quark component to be $cc\bar{u}\bar{d}$. Since the mass of the $T_{cc}^+$ is very close to the $D^*D$ threshold, it could well be interpreted as a $D^*D$ molecular state~\cite{Li:2021zbw,Ren:2021dsi,Chen:2021vhg,Dai:2021vgf,Feijoo:2021ppq,Wu:2021kbu} as predicted in many previous works~ \cite{Molina:2010tx,Li:2012ss,Liu:2019stu,Xu:2017tsr}. As early as in the 1980s, the likely existence of stable tetraquark states has attracted the interests of theorists~\cite{Ader:1981db,Ballot:1983iv,Lipkin:1986dw,Zouzou:1986qh,Heller:1986bt,Carlson:1987hh}. Latter, various models with different quark-quark interactions were employed to study the mass spectrum of  tetraquark states with the $QQ\bar{q}\bar{q}$ configuration~\cite{Manohar:1992nd,Pepin:1996id,Vijande:2006jf,Ebert:2007rn,Vijande:2013qr,Karliner:2017qjm,Eichten:2017ffp,Richard:2018yrm,Park:2018wjk,Hernandez:2019eox}. It should be noted that the $T_{cc}^+$ is the first observed double-charm exotic state. It is interesting to note that the decay width obtained from the Breit-Wigner fit and that derived from the pole position are quite different. The latter strongly supports its nature being a hadronic molecule of $DD^*$, as stressed, e.g., in Ref.~\cite{Ling:2021bir}.

The single-charm, hidden-charm, and double-charm molecular candidates have been established in experiments. In Fig.~\ref{moleculartype}, we choose the $D_{s0}^*(2317)$, $X(3872)$, and $T_{cc}^+$ states as examples and present the corresponding possible substructure. However, until now, there was no signal of triple-charm molecular states. In the future, experimental searches for triple-charm molecular states will be an interesting topic in exploring exotic hadrons.

In this work, we investigate the likely existences of triple-charm molecular states composed of $D^*D^*D$ and $D^*D^*D^*$. There are three reasons for studying such systems. First, we notice that single- and double-charm molecular candidates in Fig.~\ref{moleculartype} contain one and two charmed mesons, respectively. Thus it is  natural to ask whether there exist hadronic molecular states composed of three charmed mesons. Second, the observation of the $T_{cc}^+$ state provides a way to fix the interaction between two charmed mesons. In Ref.~\cite{Wu:2021kbu}, we  successfully reproduced the binding energy of the $T_{cc}^+$ state, with the $DD^*$ interaction  provided by the one-boson-exchange (OBE) model. This makes the numerical results more reliable when dealing with systems containing more charmed mesons in the following study. Third, in Ref.~\cite{Wu:2021kbu}, we have studied the $DDD^*$ system and found that it has a $I(J^P)=\frac{1}{2}(1^-)$ bound state solution. Compared with the $DDD^*$ system, the $D^*D^*D$ and $D^*D^*D^*$ systems can have  more spin configurations. Therefore, it is likely that there exist more hadronic molecular states in the $D^*D^*D$ and $D^*D^*D^*$ systems.

In the past several years, the LHCb Collaboration has achieved great success in discovering exotic states, including several $P_c$ states~\cite{LHCb:2015yax,LHCb:2019kea}, $P_{cs}$~\cite{LHCb:2020jpq}, $X_{0,1}(2900)$~\cite{LHCb:2020bls,LHCb:2020pxc}, and $X(6900)$~\cite{LHCb:2020bwg}. These observations demonstrated the capacity of the LHCb detector in searching for exotic states. With the upgrade of the LHCb detector~\cite{LHCb:2018roe}, one can expect that more exotic states will be observed in the future. The predictions of molecular states with the configurations of $D^*D^*D$ and $D^*D^*D^*$ may inspire more experimental works along this line.

This paper is organized as follows. In Sec.~\ref{formulism}, we introduce the interactions between  $D^*$ and $D^{(*)}$ and present the details of the Gussian expansion method. Next in Sec.~\ref{numericalresults}, we present the binding  energies and root-mean-square radii  of the $D^*D^*D$ and $D^*D^*D^*$ systems. Finally, this paper ends with a short summary in Sec.~\ref{summary}.

\section{formalism}\label{formulism}

\begin{figure}[htbp]
\includegraphics[width=6cm]{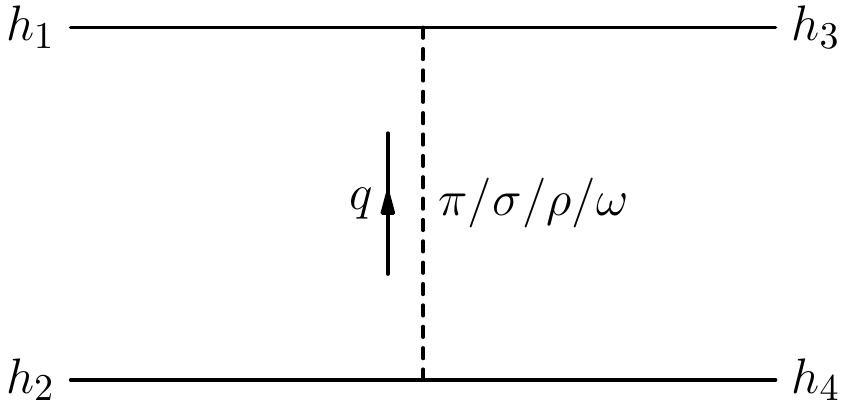}
\caption{The Feynman diagrams for the $h_1h_2\to h_3h_4$ process. In this work, the $h_1$, $h_2$, $h_3$, and $h_4$ are $D^{(*)}$ mesons.}
\label{scattering}
\end{figure}

In order to study the $D^*D^*D$ and $D^*D^*D^*$ systems, we should first derive the effective potentials of the $D^*$-$D^*$ and $D$-$D^*$ pairs. For this purpose, we adopt  the OBE model of Ref.~\cite{Liu:2019stu}. In the OBE model, the $D^*$-$D^{(*)}$ interactions occur by the scattering process as shown in Fig.~\ref{scattering}, where we should consider the exchanges of $\pi$, $\sigma$, $\rho$, and $\omega$ mesons. Then, in the momentum space, the effective potential related to the scatting amplitude can be written as
\begin{equation}
V^{h_1h_2\to h_3h_4}({\bf q})=-\frac{{\cal M}^{h_1h_2\to h_3h_4}({\bf q})}{\sqrt{\prod_i2m_i\prod_i2m_f}},
\end{equation}
where ${\cal M}^{h_1h_2\to h_3h_4}({\bf q})$ is the scattering amplitude. The $m_i$ and $m_f$ are masses of the initial and final states. To take into the finite size  of the exchanged mesons, a monopole  form factor is introduced
\begin{equation}\label{formfactor}
{\cal F}(q^2,m_E^2)=
\frac{\Lambda^2-m_E^2}{\Lambda^2-q^2},
\end{equation}
where  $q$ and $m_E$ are the mass and momentum of the exchanged meson, respectively. The effective potentials in the coordinate space can be obtained by the following Fourier transformation
\begin{equation}
V^{h_1h_2\to h_3h_4}({\bf r})=\int\frac{{\rm d}^3{\bf q}}{(2\pi)^3}{\rm e}^{i{\bf q}\cdot{\bf r}}V^{h_1h_2\to h_3h_4}({\bf q}){\cal F}^2(q^2,m_E^2).
\end{equation}
In the following, we present the effective potentials of the $D^*$-$D^{(*)}$ interactions explicitly, i.e.,
\begin{equation}\label{potential}
\begin{split}
V^{DD^*\to DD^*}=&-g_\sigma^2{\cal O}_1Y_\sigma+\frac{1}{2}\beta^2g_V^2{\cal O}_1\left({\cal C}_1(I)Y_\rho+{\cal C}_0(I)Y_\omega\right),\\
V^{DD^*\to D^*D}=&\frac{g^2}{3f_\pi^2}({\cal O}_2\hat{\cal P}+{\cal O}_3\hat{\cal Q}){\cal C}_1^\prime(I)Y_{\pi 1}\\
&+\frac{2}{3}\lambda^2g_V^2(2{\cal O}_2\hat{\cal P}-{\cal O}_3\hat{\cal Q})\left({\cal C}_1^\prime(I) Y_{\rho 1}+{\cal C}_0^\prime(I)Y_{\omega 1}\right),\\
V^{DD^*\to D^*D^*}=&\frac{g^2}{3f_\pi^2}({\cal O}_4\hat{\cal P}+{\cal O}_5\hat{\cal Q}){\cal C}_1(I)Y_{\pi 2}\\
&+\frac{2}{3}\lambda^2g_V^2(2{\cal O}_4\hat{\cal P}-{\cal O}_5\hat{\cal Q})\left({\cal C}_1(I) Y_{\rho 2}+{\cal C}_0(I)Y_{\omega 2}\right),\\
V^{D^*D^*\to D^*D^*}=&-g_\sigma^2{\cal O}_6Y_\sigma+\frac{1}{2}\beta^2g_V^2{\cal O}_6\left({\cal C}_1(I)Y_\rho+{\cal C}_0(I)Y_\omega\right)\\
        &-\frac{g^2}{3f_\pi^2}({\cal O}_7\hat{\cal P}+{\cal O}_8\hat{\cal Q}){\cal C}_1(I)Y_{\pi}\\
&-\frac{2}{3}\lambda^2g_V^2(2{\cal O}_7\hat{\cal P}-{\cal O}_8\hat{\cal Q})\left({\cal C}_1(I) Y_{\rho}+{\cal C}_0(I)Y_{\omega}\right).\\
\end{split}
\end{equation}
In Eq.~(\ref{potential}), the ${\cal O}_i$'s are spin-dependent operators, which are defined as
\begin{equation}
\begin{split}
{\cal O}_1=&{\bm\epsilon}_4^\dagger\cdot{\bm\epsilon}_2,\\
{\cal O}_2=&{\bm\epsilon}_3^\dagger\cdot{\bm\epsilon}_2,\\
{\cal O}_3=&S({\bf r},{\bm \epsilon}_3^\dagger,{\bm \epsilon}_2),\\
{\cal O}_4=&{\bm\epsilon}_3^\dagger\cdot(i{\bm\epsilon}_4^\dagger\times{\bm\epsilon}_2),\\
{\cal O}_5=&S({\bf r},{\bm\epsilon}_3^\dagger,i{\bm\epsilon}_4^\dagger\times{\bm\epsilon}_2),\\
{\cal O}_6=&({\bm\epsilon}_3^\dagger\cdot{\bm\epsilon}_1)({\bm\epsilon}_4^\dagger\cdot{\bm\epsilon}_2),\\
{\cal O}_7=&({\bm\epsilon}_3^\dagger\times{\bm\epsilon}_1)\cdot({\bm\epsilon}_4^\dagger\times{\bm\epsilon}_2),\\
{\cal O}_8=&S({\bf r},{\bm\epsilon}_3^\dagger\times{\bm\epsilon}_1,{\bm\epsilon}_4^\dagger\times{\bm\epsilon}_2),\\
\end{split}
\end{equation}
where
\begin{equation}
S({\bf r},{\bf a},{\bf b})=\frac{3({\bf a}\cdot {\bf r})({\bf b}\cdot {\bf r})}{r^2}-{\bf a}\cdot{\bf b}
\end{equation}
is the tensor operator. Here,  ${\bm\epsilon}_{i}$ ($i=1,2$) and ${\bm\epsilon}^\dagger_{i}$ ($i=3,4$) are initial and final polarization vectors of the $D^*$ mesons, respectively, and $C_{0}^{(\prime)}(I)$ and $C_{1}^{(\prime)}(I)$ are flavor-dependent factors given by
\begin{equation}
\begin{split}
&{\cal C}_1(0)=-\frac{3}{2},~{\cal C}_1^\prime(0)=+\frac{3}{2},~{\cal C}_0(0)=+\frac{1}{2},~{\cal C}_0^\prime(0)=-\frac{1}{2},\\
&{\cal C}_1(1)=+\frac{1}{2},~{\cal C}_1^\prime(1)=+\frac{1}{2},~{\cal C}_0(1)=+\frac{1}{2},~{\cal C}_0^\prime(1)=+\frac{1}{2}.
\end{split}
\end{equation}
The function $Y_i$ in Eq.~(\ref{potential}) is written as\footnote{
In momentum space, the effective potentials share a common part, i.e.,
\begin{equation}\label{Vqcommon}
V({\bf q})=\frac{1}{{\bf q}^2+m_E^2-{q^0}^2},
\end{equation}
where the $q^0$ is energy component of the exchange momentum, whose explicit expression could be found in Ref.~\cite{Li:2012ss}. Without a form factor, the Fourier transformation of Eq.~(\ref{Vqcommon}) is
\begin{equation}
\begin{split}
V({\bf r})=&\int\frac{{\rm d}^3{\bf q}}{(2\pi)^3}{\rm e}^{i{\bf q}\cdot{\bf r}}\frac{1}{{\bf q}^2+m_E^2-{q^0}^2}\\
&=\frac{1}{4\pi r}{\rm e}^{-\sqrt{m_E^2-{q^0}^2}r}.
\end{split}
\end{equation}
After introducing the form factor, the Fourier transformation is
\begin{equation}
\begin{split}
V({\bf r})=&\int\frac{{\rm d}^3{\bf q}}{(2\pi)^3}{\rm e}^{i{\bf q}\cdot{\bf r}}\frac{1}{{\bf q}^2+m_E^2-{q^0}^2}{\cal F}^2(q^2,m_E^2).
\end{split}
\end{equation}
After performing the integration, one could obtain the function $Y_i$ given in Eq.~(\ref{YiFunction}).
}
\begin{equation}\label{YiFunction}
Y_i=\frac{{\rm e}^{-m_{Ei}r}}{4\pi r}-\frac{{\rm e}^{-\Lambda_i r}}{4\pi r}-\frac{\Lambda_i^2{\rm e}^{-\Lambda_i r}}{8\pi\Lambda_i}+\frac{m_{Ei}^2{\rm e}^{-\Lambda_i r}}{8\pi\Lambda_i}
\end{equation}
with $\Lambda_i=\sqrt{\Lambda^2-q_i^2}$ and $m_{Ei}=\sqrt{m_E^2-q_i^2}$. The $q_i$ is the energy component of the exchanged momentum. We employ $q_1=m_{D^*}-m_D$
and $q_2=(m_{D^*}^2-m_D^2)/(4m_{D^*})$ for the $DD^*\to D^*D$ and $DD^*\to D^*D^*$ processes, respectively. The operators $\hat{\cal P}$ and $\hat{\cal Q}$ only act on  $Y_i$  and the expressions are
\begin{equation}
\hat{\cal P}=\frac{1}{r^2}\frac{\partial}{\partial r}r^2\frac{\partial}{\partial r},~~~
\hat{\cal Q}=r\frac{\partial}{\partial r}\frac{1}{r}\frac{\partial}{\partial r}.
\end{equation}
To evaluate the above potentials, we also need the values of the coupling constants and the masses of the mesons, which are collected in Table~\ref{parameters}.

\begin{table}[htbp]
\centering
\renewcommand\arraystretch{1.20}
\caption{Values of the coupling constants~\cite{Liu:2019stu,Riska:2000gd,Gell-Mann:1960mvl,Isola:2003fh} and meson masses~\cite{ParticleDataGroup:2020ssz}. }
\label{parameters}
\begin{tabular*}{86mm}{@{\extracolsep{\fill}}cc@{\hskip\tabcolsep\vrule width 0.75pt\hskip\tabcolsep}cc}
\toprule[1.00pt]
\toprule[1.00pt]
Coupling Constants &Values           &Mesons   &Mass (GeV)\\
\midrule[0.75pt]
$g$                &0.6              &$\pi$    &0.140     \\
$f_\pi$            &0.132 GeV        &$\sigma$ &0.600     \\
$g_\sigma$         &3.4              &$\rho$   &0.770     \\
$\beta g_V$        &5.2              &$\omega$ &0.780     \\
$\lambda g_V$      &3.133 GeV$^{-1}$ &$D$      &1.867     \\
                   &                 &$D^*$    &2.009     \\
\bottomrule[1pt]
\bottomrule[1pt]
\end{tabular*}
\end{table}

\begin{figure}[t]
\includegraphics[width=8.6cm]{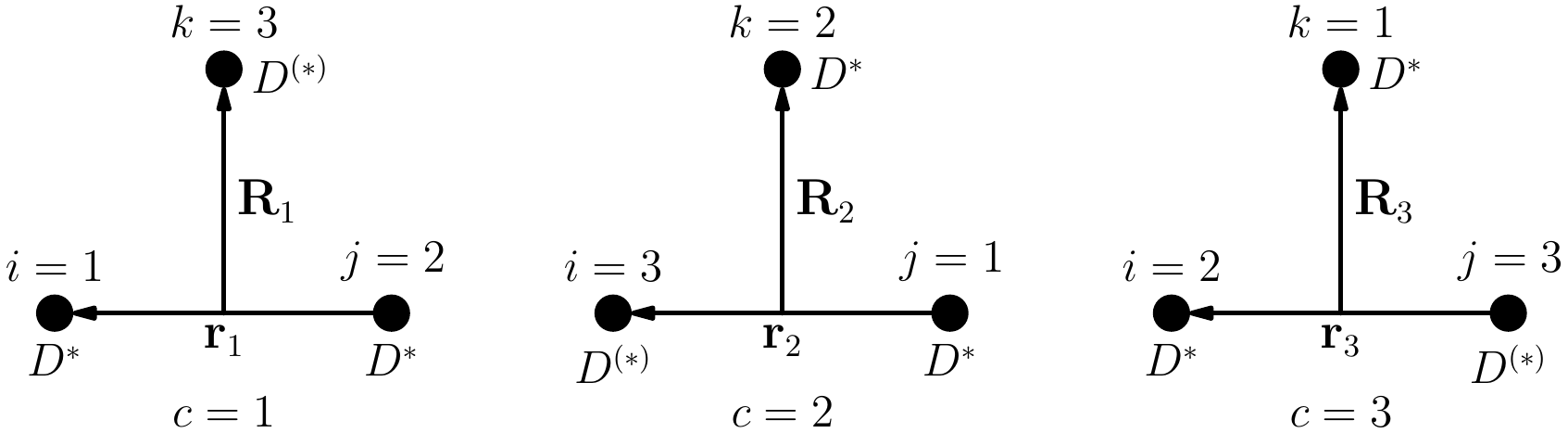}
\caption{Jacobi coordinates of the $D^*D^*D^{(*)}$ systems.}
\label{Jacobi}
\end{figure}

To solve the three-body Schr\"odinger equation, we employ the Gaussian expansion method~\cite{Hiyama:2003cu,Hiyama:2012sma}, which was widely used in studies of baryon systems~\cite{Yoshida:2015tia,Yang:2019lsg,Yang:2017qan}, multiquark states~\cite{Yang:2021izl,Yang:2020fou,Wang:2019rdo,Lu:2021kut,Lu:2020cns}, and multibody hadronic molecular states~\cite{Wu:2019vsy,Wu:2020rdg,Wu:2020job,Wu:2021kbu,Wu:2021gyn} (for reviews on this latter topic, see, e.g., Refs.~\cite{Wu:2021ljz,Wu:2021dwy}). The three-body Schr\"odinger equation reads
\begin{equation}\label{threebodySch}
\left[\hat{T}+V(r_1)+V(r_2)+V(r_3)\right]\Psi_{JM}=E\Psi_{JM},
\end{equation}
where  $\hat{T}$ is the kinetic energy operator, and $V(r_1)$, $V(r_2)$, and $V(r_3)$ are pairwise potentials. $\Psi_{JM}$ is the total wave function, which can be written as
\begin{equation}\label{PsiJM}
\Psi_{JM}=\sum_{c,\alpha}C_{c,\alpha} \Psi_{JM}^{(c,\alpha)},
\end{equation}
where
\begin{equation}
\begin{split}
\Psi_{JM}^{(c,\alpha)}=H_{t,T}^c\left[\chi_{s,S}^c\left[\phi_{nl}({\bf r}_c)\phi_{NL}({\bf R}_c)\right]_\lambda\right]_{JM}
\end{split}
\end{equation}
is the basis, and $C_{c,\alpha}$ is the coefficient of the corresponding basis, which can be obtained by the Rayleigh-Ritz variational method. The $c$ ($c=1,2,3$) represents the three channels in Fig.~\ref{Jacobi} and $\alpha=\{tT,sS,nN,lL\lambda\}$ is the quantum number  of the basis.  $H_{t,T}^c$ is the flavor wave function, where  $t$ is isospin in the ${\bf r}_c$ degree of freedom and $T$ is the total isospin. $\chi_{s,S}^c$ is the spin wave function, where the $s$, $S$ are spin in the ${\bf r}_c$ degree of freedom and total spin, respectively. $\phi_{nlm_l}({\bf r}_c)$ and $\phi_{NLM_L}({\bf R}_c)$ are spatial wave functions, which read
\begin{equation}\label{gaussfun}
\begin{split}
\phi_{nlm_l}({\bf r}_c)=&N_{nl}r_c^le^{-\nu_nr_c^2}Y_{lm}(\hat{\bf r}_c),\\
\phi_{NLM_L}({\bf R}_c)=&N_{NL}R_c^Le^{-\lambda_NR_c^2}Y_{LM}(\hat{\bf R}_c),\\
\end{split}
\end{equation}
where $N_{nl}$ and $N_{NL}$ are normalization constants. In Eq.~(\ref{gaussfun}), The ${\bf r}_c$ and ${\bf R}_c$ are Jacobi coordinates, and $\nu_n$ and $\lambda_N$  are Gaussian ranges. After the above preparations, we can calculate the kinetic, potential, and normalization matrix elements (see Refs.~\cite{Hiyama:2003cu,Brink:1998as} for more details), i.e.,
\begin{equation}
\begin{split}
T^{ab}_{\alpha\alpha^\prime}=&\langle\Psi_{JM}^{(a,\alpha)}|\hat{T}|\Psi_{JM}^{(b,\alpha^\prime)}\rangle,\\
V^{ab}_{\alpha\alpha^\prime c}=&\langle\Psi_{JM}^{(a,\alpha)}|V({r_c})|\Psi_{JM}^{(b,\alpha^\prime)}\rangle,\\
N^{ab}_{\alpha\alpha^\prime}=&\langle\Psi_{JM}^{(a,\alpha)}|\Psi_{JM}^{(b,\alpha^\prime)}\rangle.
\end{split}
\end{equation}
Then, Eq.~(\ref{threebodySch}) could be further expressed as the following general eigenvalue equation:
\begin{equation}
\left(T^{ab}_{\alpha\alpha^\prime}+\sum\limits_{c=1}^3V^{ab}_{\alpha\alpha^\prime c}\right)C_{b,\alpha^\prime}=EN^{ab}_{\alpha\alpha^\prime}C_{b,\alpha^\prime}.
\end{equation}

\section{numerical results}\label{numericalresults}

\begin{table*}
\caption{Configurations of the $D^*D^*D$-$D^*D^*D^*$ systems. The $R_{lL\lambda}^c=\left[\phi_{nl}({\bf r}_c)\phi_{NL}({\bf R}_c)\right]_\lambda$, $\chi_{sS}^c$, and $H_{tT}^c$ represent the spatial, spin, and flavor wave functions, respectively.}
\label{bases}
\centering
\renewcommand\arraystretch{1.4}
\begin{tabular}{m{0.3cm}<{\centering}m{0.4cm}<{\centering}@{\vrule width 0.75pt}m{3.5cm}@{\vrule width 0.75pt}m{6.4cm}@{\vrule width 0.75pt}m{7cm}}
\toprule[1.00pt]
\toprule[1.00pt]
\multirow{2}{*}{$I$}&\multirow{2}{*}{$J^P$}&\multicolumn{2}{c@{\vrule width 0.75pt}}{$D^*D^*D$}&\multicolumn{1}{c}{$D^*D^*D^*$}\\
\Xcline{3-5}{0.75pt}
&&\multicolumn{1}{c@{\vrule width 0.75pt}}{$c=1$}&\multicolumn{1}{c@{\vrule width 0.75pt}}{$c=2,~3$}&\multicolumn{1}{c}{$c=1,~2,~3$}\\
\midrule[0.75pt]
\multirow{4}[45]{*}{$\frac{1}{2}$}
&$0^-$
&\scalebox{0.8}{\makecell[l]{
$R_{202}^c\chi_{2,2}^c H_{1,\frac{1}{2}}^c$,~$R_{022}^c\chi_{2,2}^c H_{1,\frac{1}{2}}^c$,~\\
$R_{112}^c\chi_{2,2}^c H_{0,\frac{1}{2}}^c$,~$R_{000}^c\chi_{0,0}^c H_{1,\frac{1}{2}}^c$,~\\
$R_{111}^c\chi_{1,1}^c H_{1,\frac{1}{2}}^c$,~$R_{110}^c\chi_{0,0}^c H_{0,\frac{1}{2}}^c$\\
}}
&\scalebox{0.8}{\makecell[l]{
$R_{202}^c \{\chi_{1,2}^c H_{0,\frac{1}{2}}^c,\chi_{1,2}^c H_{1,\frac{1}{2}}^c\}$,~$R_{022}^c \{\chi_{1,2}^c H_{0,\frac{1}{2}}^c,\chi_{1,2}^c H_{1,\frac{1}{2}}^c\}$,~\\
$R_{112}^c \{\chi_{1,2}^c H_{0,\frac{1}{2}}^c,\chi_{1,2}^c H_{1,\frac{1}{2}}^c\}$,~$R_{000}^c \{\chi_{1,0}^c H_{0,\frac{1}{2}}^c,\chi_{1,0}^c H_{1,\frac{1}{2}}^c\}$,~\\
$R_{111}^c \{\chi_{1,1}^c H_{0,\frac{1}{2}}^c,\chi_{1,1}^c H_{1,\frac{1}{2}}^c\}$,~$R_{110}^c \{\chi_{1,0}^c H_{0,\frac{1}{2}}^c,\chi_{1,0}^c H_{1,\frac{1}{2}}^c\}$\\
}}
&\scalebox{0.8}{\makecell[l]{
$R_{000}^c\chi_{1,0}^c H_{0,\frac{1}{2}}^c$,~$R_{110}^c\chi_{1,0}^c H_{1,\frac{1}{2}}^c$,~$R_{202}^c \{\chi_{1,2}^c H_{0,\frac{1}{2}}^c,\chi_{2,2}^c H_{1,\frac{1}{2}}^c\}$,~\\
$R_{022}^c \{\chi_{1,2}^c H_{0,\frac{1}{2}}^c,\chi_{2,2}^c H_{1,\frac{1}{2}}^c\}$,~$R_{112}^c \{\chi_{1,2}^c H_{1,\frac{1}{2}}^c,\chi_{2,2}^c H_{0,\frac{1}{2}}^c\}$,~\\
$R_{111}^c \{\chi_{0,1}^c H_{0,\frac{1}{2}}^c,\chi_{1,1}^c H_{1,\frac{1}{2}}^c,\chi_{2,1}^c H_{0,\frac{1}{2}}^c\}$\\
}}\\
\Xcline{2-5}{0.75pt}
&$1^-$
&\scalebox{0.8}{\makecell[l]{
$R_{000}^c\chi_{1,1}^c H_{0,\frac{1}{2}}^c$,~$R_{110}^c\chi_{1,1}^c H_{1,\frac{1}{2}}^c$,~\\
$R_{202}^c \{\chi_{1,1}^c H_{0,\frac{1}{2}}^c,\chi_{2,2}^c H_{1,\frac{1}{2}}^c\}$,~\\
$R_{022}^c \{\chi_{1,1}^c H_{0,\frac{1}{2}}^c,\chi_{2,2}^c H_{1,\frac{1}{2}}^c\}$,~\\
$R_{112}^c \{\chi_{1,1}^c H_{1,\frac{1}{2}}^c,\chi_{2,2}^c H_{0,\frac{1}{2}}^c\}$,~\\
$R_{111}^c \{\chi_{0,0}^c H_{0,\frac{1}{2}}^c,\chi_{1,1}^c H_{1,\frac{1}{2}}^c,\chi_{2,2}^c H_{0,\frac{1}{2}}^c\}$\\
}}
&\scalebox{0.8}{\makecell[l]{
$R_{000}^c \{\chi_{1,1}^c H_{0,\frac{1}{2}}^c,\chi_{1,1}^c H_{1,\frac{1}{2}}^c\}$,~$R_{110}^c \{\chi_{1,1}^c H_{0,\frac{1}{2}}^c,\chi_{1,1}^c H_{1,\frac{1}{2}}^c\}$,~\\
$R_{202}^c \{\chi_{1,1}^c H_{0,\frac{1}{2}}^c,\chi_{1,2}^c H_{0,\frac{1}{2}}^c,\chi_{1,1}^c H_{1,\frac{1}{2}}^c,\chi_{1,2}^c H_{1,\frac{1}{2}}^c\}$,~\\
$R_{022}^c \{\chi_{1,1}^c H_{0,\frac{1}{2}}^c,\chi_{1,2}^c H_{0,\frac{1}{2}}^c,\chi_{1,1}^c H_{1,\frac{1}{2}}^c,\chi_{1,2}^c H_{1,\frac{1}{2}}^c\}$,~\\
$R_{112}^c \{\chi_{1,1}^c H_{0,\frac{1}{2}}^c,\chi_{1,2}^c H_{0,\frac{1}{2}}^c,\chi_{1,1}^c H_{1,\frac{1}{2}}^c,\chi_{1,2}^c H_{1,\frac{1}{2}}^c\}$,~\\
$R_{111}^c \{\chi_{1,0}^c H_{0,\frac{1}{2}}^c,\chi_{1,1}^c H_{0,\frac{1}{2}}^c,\chi_{1,2}^c H_{0,\frac{1}{2}}^c,\chi_{1,0}^c H_{1,\frac{1}{2}}^c,\chi_{1,1}^c H_{1,\frac{1}{2}}^c,\chi_{1,2}^c H_{1,\frac{1}{2}}^c\}$\\
}}
&\scalebox{0.8}{\makecell[l]{
$R_{000}^c \{\chi_{0,1}^c H_{1,\frac{1}{2}}^c,\chi_{1,1}^c H_{0,\frac{1}{2}}^c,\chi_{2,1}^c H_{1,\frac{1}{2}}^c\}$,~\\
$R_{110}^c \{\chi_{0,1}^c H_{0,\frac{1}{2}}^c,\chi_{1,1}^c H_{1,\frac{1}{2}}^c,\chi_{2,1}^c H_{0,\frac{1}{2}}^c\}$,~\\
$R_{202}^c \{\chi_{0,1}^c H_{1,\frac{1}{2}}^c,\chi_{1,1}^c H_{0,\frac{1}{2}}^c,\chi_{1,2}^c H_{0,\frac{1}{2}}^c,\chi_{2,1}^c H_{1,\frac{1}{2}}^c,\chi_{2,2}^c H_{1,\frac{1}{2}}^c,\chi_{2,3}^c H_{1,\frac{1}{2}}^c\}$,~\\
$R_{022}^c \{\chi_{0,1}^c H_{1,\frac{1}{2}}^c,\chi_{1,1}^c H_{0,\frac{1}{2}}^c,\chi_{1,2}^c H_{0,\frac{1}{2}}^c,\chi_{2,1}^c H_{1,\frac{1}{2}}^c,\chi_{2,2}^c H_{1,\frac{1}{2}}^c,\chi_{2,3}^c H_{1,\frac{1}{2}}^c\}$,~\\
$R_{112}^c \{\chi_{0,1}^c H_{0,\frac{1}{2}}^c,\chi_{1,1}^c H_{1,\frac{1}{2}}^c,\chi_{1,2}^c H_{1,\frac{1}{2}}^c,\chi_{2,1}^c H_{0,\frac{1}{2}}^c,\chi_{2,2}^c H_{0,\frac{1}{2}}^c,\chi_{2,3}^c H_{0,\frac{1}{2}}^c\}$,~\\
$R_{111}^c \{\chi_{0,1}^c H_{0,\frac{1}{2}}^c,\chi_{1,0}^c H_{1,\frac{1}{2}}^c,\chi_{1,1}^c H_{1,\frac{1}{2}}^c,\chi_{1,2}^c H_{1,\frac{1}{2}}^c,\chi_{2,1}^c H_{0,\frac{1}{2}}^c,\chi_{2,2}^c H_{0,\frac{1}{2}}^c\}$\\
}}\\
\Xcline{2-5}{0.75pt}
&$2^-$
&\scalebox{0.8}{\makecell[l]{
$R_{000}^c\chi_{2,2}^c H_{1,\frac{1}{2}}^c$,~$R_{110}^c\chi_{2,2}^c H_{0,\frac{1}{2}}^c$,~\\
$R_{111}^c \{\chi_{1,1}^c H_{1,\frac{1}{2}}^c,\chi_{2,2}^c H_{0,\frac{1}{2}}^c\}$,~\\
$R_{202}^c \{\chi_{0,0}^c H_{1,\frac{1}{2}}^c,\chi_{1,1}^c H_{0,\frac{1}{2}}^c,\chi_{2,2}^c H_{1,\frac{1}{2}}^c\}$,~\\
$R_{022}^c \{\chi_{0,0}^c H_{1,\frac{1}{2}}^c,\chi_{1,1}^c H_{0,\frac{1}{2}}^c,\chi_{2,2}^c H_{1,\frac{1}{2}}^c\}$,~\\
$R_{112}^c \{\chi_{0,0}^c H_{0,\frac{1}{2}}^c,\chi_{1,1}^c H_{1,\frac{1}{2}}^c,\chi_{2,2}^c H_{0,\frac{1}{2}}^c\}$\\
}}
&\scalebox{0.8}{\makecell[l]{
$R_{000}^c \{\chi_{1,2}^c H_{0,\frac{1}{2}}^c,\chi_{1,2}^c H_{1,\frac{1}{2}}^c\}$,~$R_{110}^c \{\chi_{1,2}^c H_{0,\frac{1}{2}}^c,\chi_{1,2}^c H_{1,\frac{1}{2}}^c\}$,~\\
$R_{111}^c \{\chi_{1,1}^c H_{0,\frac{1}{2}}^c,\chi_{1,2}^c H_{0,\frac{1}{2}}^c,\chi_{1,1}^c H_{1,\frac{1}{2}}^c,\chi_{1,2}^c H_{1,\frac{1}{2}}^c\}$,~\\
$R_{202}^c \{\chi_{1,0}^c H_{0,\frac{1}{2}}^c,\chi_{1,1}^c H_{0,\frac{1}{2}}^c,\chi_{1,2}^c H_{0,\frac{1}{2}}^c,\chi_{1,0}^c H_{1,\frac{1}{2}}^c,\chi_{1,1}^c H_{1,\frac{1}{2}}^c,\chi_{1,2}^c H_{1,\frac{1}{2}}^c\}$,~\\
$R_{022}^c \{\chi_{1,0}^c H_{0,\frac{1}{2}}^c,\chi_{1,1}^c H_{0,\frac{1}{2}}^c,\chi_{1,2}^c H_{0,\frac{1}{2}}^c,\chi_{1,0}^c H_{1,\frac{1}{2}}^c,\chi_{1,1}^c H_{1,\frac{1}{2}}^c,\chi_{1,2}^c H_{1,\frac{1}{2}}^c\}$,~\\
$R_{112}^c \{\chi_{1,0}^c H_{0,\frac{1}{2}}^c,\chi_{1,1}^c H_{0,\frac{1}{2}}^c,\chi_{1,2}^c H_{0,\frac{1}{2}}^c,\chi_{1,0}^c H_{1,\frac{1}{2}}^c,\chi_{1,1}^c H_{1,\frac{1}{2}}^c,\chi_{1,2}^c H_{1,\frac{1}{2}}^c\}$\\
}}
&\scalebox{0.8}{\makecell[l]{
$R_{000}^c \{\chi_{1,2}^c H_{0,\frac{1}{2}}^c,\chi_{2,2}^c H_{1,\frac{1}{2}}^c\}$,~$R_{110}^c \{\chi_{1,2}^c H_{1,\frac{1}{2}}^c,\chi_{2,2}^c H_{0,\frac{1}{2}}^c\}$,~\\
$R_{111}^c \{\chi_{0,1}^c H_{0,\frac{1}{2}}^c,\chi_{1,1}^c H_{1,\frac{1}{2}}^c,\chi_{1,2}^c H_{1,\frac{1}{2}}^c,\chi_{2,1}^c H_{0,\frac{1}{2}}^c,\chi_{2,2}^c H_{0,\frac{1}{2}}^c,\chi_{2,3}^c H_{0,\frac{1}{2}}^c\}$,~\\
$R_{202}^c \{\chi_{0,1}^c H_{1,\frac{1}{2}}^c,\chi_{1,0}^c H_{0,\frac{1}{2}}^c,\chi_{1,1}^c H_{0,\frac{1}{2}}^c,\chi_{1,2}^c H_{0,\frac{1}{2}}^c,\chi_{2,1}^c H_{1,\frac{1}{2}}^c,\chi_{2,2}^c H_{1,\frac{1}{2}}^c,\chi_{2,3}^c H_{1,\frac{1}{2}}^c\}$,~\\
$R_{022}^c \{\chi_{0,1}^c H_{1,\frac{1}{2}}^c,\chi_{1,0}^c H_{0,\frac{1}{2}}^c,\chi_{1,1}^c H_{0,\frac{1}{2}}^c,\chi_{1,2}^c H_{0,\frac{1}{2}}^c,\chi_{2,1}^c H_{1,\frac{1}{2}}^c,\chi_{2,2}^c H_{1,\frac{1}{2}}^c,\chi_{2,3}^c H_{1,\frac{1}{2}}^c\}$,~\\
$R_{112}^c \{\chi_{0,1}^c H_{0,\frac{1}{2}}^c,\chi_{1,0}^c H_{1,\frac{1}{2}}^c,\chi_{1,1}^c H_{1,\frac{1}{2}}^c,\chi_{1,2}^c H_{1,\frac{1}{2}}^c,\chi_{2,1}^c H_{0,\frac{1}{2}}^c,\chi_{2,2}^c H_{0,\frac{1}{2}}^c,\chi_{2,3}^c H_{0,\frac{1}{2}}^c\}$\\
}}\\
\Xcline{2-5}{0.75pt}
&$3^-$
&
&
&\scalebox{0.8}{\makecell[l]{
$R_{000}^c\chi_{2,3}^c H_{1,\frac{1}{2}}^c$,~$R_{110}^c\chi_{2,3}^c H_{0,\frac{1}{2}}^c$,~$R_{111}^c \{\chi_{1,2}^c H_{1,\frac{1}{2}}^c,\chi_{2,2}^c H_{0,\frac{1}{2}}^c,\chi_{2,3}^c H_{0,\frac{1}{2}}^c\}$,~\\
$R_{202}^c \{\chi_{0,1}^c H_{1,\frac{1}{2}}^c,\chi_{1,1}^c H_{0,\frac{1}{2}}^c,\chi_{1,2}^c H_{0,\frac{1}{2}}^c,\chi_{2,1}^c H_{1,\frac{1}{2}}^c,\chi_{2,2}^c H_{1,\frac{1}{2}}^c,\chi_{2,3}^c H_{1,\frac{1}{2}}^c\}$,~\\
$R_{022}^c \{\chi_{0,1}^c H_{1,\frac{1}{2}}^c,\chi_{1,1}^c H_{0,\frac{1}{2}}^c,\chi_{1,2}^c H_{0,\frac{1}{2}}^c,\chi_{2,1}^c H_{1,\frac{1}{2}}^c,\chi_{2,2}^c H_{1,\frac{1}{2}}^c,\chi_{2,3}^c H_{1,\frac{1}{2}}^c\}$,~\\
$R_{112}^c \{\chi_{0,1}^c H_{0,\frac{1}{2}}^c,\chi_{1,1}^c H_{1,\frac{1}{2}}^c,\chi_{1,2}^c H_{1,\frac{1}{2}}^c,\chi_{2,1}^c H_{0,\frac{1}{2}}^c,\chi_{2,2}^c H_{0,\frac{1}{2}}^c,\chi_{2,3}^c H_{0,\frac{1}{2}}^c\}$\\
}}\\
\midrule[0.75pt]
\multirow{4}[35]{*}{$\frac{3}{2}$}
&$0^-$
&\scalebox{0.8}{\makecell[l]{
$R_{202}^c\chi_{2,2}^c H_{1,\frac{3}{2}}^c$,~$R_{000}^c\chi_{0,0}^c H_{1,\frac{3}{2}}^c$,~\\
$R_{022}^c\chi_{2,2}^c H_{1,\frac{3}{2}}^c$,~$R_{111}^c\chi_{1,1}^c H_{1,\frac{3}{2}}^c$\\
}}
&\scalebox{0.8}{\makecell[l]{
$R_{202}^c\chi_{1,2}^c H_{1,\frac{3}{2}}^c$,~$R_{022}^c\chi_{1,2}^c H_{1,\frac{3}{2}}^c$,~$R_{112}^c\chi_{1,2}^c H_{1,\frac{3}{2}}^c$,~\\
$R_{000}^c\chi_{1,0}^c H_{1,\frac{3}{2}}^c$,~$R_{111}^c\chi_{1,1}^c H_{1,\frac{3}{2}}^c$,~$R_{110}^c\chi_{1,0}^c H_{1,\frac{3}{2}}^c$\\
}}
&\scalebox{0.8}{\makecell[l]{
$R_{202}^c\chi_{2,2}^c H_{1,\frac{3}{2}}^c$,~$R_{022}^c\chi_{2,2}^c H_{1,\frac{3}{2}}^c$,~$R_{112}^c\chi_{1,2}^c H_{1,\frac{3}{2}}^c$,~\\
$R_{111}^c\chi_{1,1}^c H_{1,\frac{3}{2}}^c$,~$R_{110}^c\chi_{1,0}^c H_{1,\frac{3}{2}}^c$}}\\
\Xcline{2-5}{0.75pt}
&$1^-$
&\scalebox{0.8}{\makecell[l]{
$R_{202}^c\chi_{2,2}^c H_{1,\frac{3}{2}}^c$,~$R_{022}^c\chi_{2,2}^c H_{1,\frac{3}{2}}^c$,~\\
$R_{112}^c\chi_{1,1}^c H_{1,\frac{3}{2}}^c$,~$R_{111}^c\chi_{1,1}^c H_{1,\frac{3}{2}}^c$,~\\
$R_{110}^c\chi_{1,1}^c H_{1,\frac{3}{2}}^c$}}
&\scalebox{0.8}{\makecell[l]{
$R_{000}^c\chi_{1,1}^c H_{1,\frac{3}{2}}^c$,~$R_{110}^c\chi_{1,1}^c H_{1,\frac{3}{2}}^c$,~$R_{202}^c \{\chi_{1,1}^c H_{1,\frac{3}{2}}^c,\chi_{1,2}^c H_{1,\frac{3}{2}}^c\}$,~\\
$R_{022}^c \{\chi_{1,1}^c H_{1,\frac{3}{2}}^c,\chi_{1,2}^c H_{1,\frac{3}{2}}^c\}$,~$R_{112}^c \{\chi_{1,1}^c H_{1,\frac{3}{2}}^c,\chi_{1,2}^c H_{1,\frac{3}{2}}^c\}$,~\\
$R_{111}^c \{\chi_{1,0}^c H_{1,\frac{3}{2}}^c,\chi_{1,1}^c H_{1,\frac{3}{2}}^c,\chi_{1,2}^c H_{1,\frac{3}{2}}^c\}$\\
}}
&\scalebox{0.8}{\makecell[l]{
$R_{000}^c \{\chi_{0,1}^c H_{1,\frac{3}{2}}^c,\chi_{2,1}^c H_{1,\frac{3}{2}}^c\}$,~$R_{110}^c\chi_{1,1}^c H_{1,\frac{3}{2}}^c$,~\\
$R_{112}^c \{\chi_{1,1}^c H_{1,\frac{3}{2}}^c,\chi_{1,2}^c H_{1,\frac{3}{2}}^c\}$,~$R_{111}^c \{\chi_{1,0}^c H_{1,\frac{3}{2}}^c,\chi_{1,1}^c H_{1,\frac{3}{2}}^c,\chi_{1,2}^c H_{1,\frac{3}{2}}^c\}$,~\\
$R_{202}^c \{\chi_{0,1}^c H_{1,\frac{3}{2}}^c,\chi_{2,1}^c H_{1,\frac{3}{2}}^c,\chi_{2,2}^c H_{1,\frac{3}{2}}^c,\chi_{2,3}^c H_{1,\frac{3}{2}}^c\}$,~\\
$R_{022}^c \{\chi_{0,1}^c H_{1,\frac{3}{2}}^c,\chi_{2,1}^c H_{1,\frac{3}{2}}^c,\chi_{2,2}^c H_{1,\frac{3}{2}}^c,\chi_{2,3}^c H_{1,\frac{3}{2}}^c\}$\\
}}\\
\Xcline{2-5}{0.75pt}
&$2^-$
&\scalebox{0.8}{\makecell[l]{
$R_{000}^c\chi_{2,2}^c H_{1,\frac{3}{2}}^c$,~\\
$R_{112}^c\chi_{1,1}^c H_{1,\frac{3}{2}}^c$,~$R_{111}^c\chi_{1,1}^c H_{1,\frac{3}{2}}^c$,~\\
$R_{202}^c \{\chi_{0,0}^c H_{1,\frac{3}{2}}^c,\chi_{2,2}^c H_{1,\frac{3}{2}}^c\}$,~\\
$R_{022}^c \{\chi_{0,0}^c H_{1,\frac{3}{2}}^c,\chi_{2,2}^c H_{1,\frac{3}{2}}^c\}$\\
}}
&\scalebox{0.8}{\makecell[l]{
$R_{000}^c\chi_{1,2}^c H_{1,\frac{3}{2}}^c$,~$R_{110}^c\chi_{1,2}^c H_{1,\frac{3}{2}}^c$,~$R_{111}^c \{\chi_{1,1}^c H_{1,\frac{3}{2}}^c,\chi_{1,2}^c H_{1,\frac{3}{2}}^c\}$,~\\
$R_{202}^c \{\chi_{1,0}^c H_{1,\frac{3}{2}}^c,\chi_{1,1}^c H_{1,\frac{3}{2}}^c,\chi_{1,2}^c H_{1,\frac{3}{2}}^c\}$,~\\
$R_{022}^c \{\chi_{1,0}^c H_{1,\frac{3}{2}}^c,\chi_{1,1}^c H_{1,\frac{3}{2}}^c,\chi_{1,2}^c H_{1,\frac{3}{2}}^c\}$,~\\
$R_{112}^c \{\chi_{1,0}^c H_{1,\frac{3}{2}}^c,\chi_{1,1}^c H_{1,\frac{3}{2}}^c,\chi_{1,2}^c H_{1,\frac{3}{2}}^c\}$\\
}}
&\scalebox{0.8}{\makecell[l]{
$R_{000}^c\chi_{2,2}^c H_{1,\frac{3}{2}}^c$,~$R_{110}^c\chi_{1,2}^c H_{1,\frac{3}{2}}^c$,~$R_{111}^c \{\chi_{1,1}^c H_{1,\frac{3}{2}}^c,\chi_{1,2}^c H_{1,\frac{3}{2}}^c\}$,~\\
$R_{112}^c \{\chi_{1,0}^c H_{1,\frac{3}{2}}^c,\chi_{1,1}^c H_{1,\frac{3}{2}}^c,\chi_{1,2}^c H_{1,\frac{3}{2}}^c\}$,~\\
$R_{202}^c \{\chi_{0,1}^c H_{1,\frac{3}{2}}^c,\chi_{2,1}^c H_{1,\frac{3}{2}}^c,\chi_{2,2}^c H_{1,\frac{3}{2}}^c,\chi_{2,3}^c H_{1,\frac{3}{2}}^c\}$,~\\
$R_{022}^c \{\chi_{0,1}^c H_{1,\frac{3}{2}}^c,\chi_{2,1}^c H_{1,\frac{3}{2}}^c,\chi_{2,2}^c H_{1,\frac{3}{2}}^c,\chi_{2,3}^c H_{1,\frac{3}{2}}^c\}$\\
}}\\
\Xcline{2-5}{0.75pt}
&$3^-$
&
&
&\scalebox{0.8}{\makecell[l]{
$R_{000}^c\chi_{2,3}^c H_{1,\frac{3}{2}}^c$,~$R_{111}^c\chi_{1,2}^c H_{1,\frac{3}{2}}^c$,~$R_{112}^c \{\chi_{1,1}^c H_{1,\frac{3}{2}}^c,\chi_{1,2}^c H_{1,\frac{3}{2}}^c\}$,~\\
$R_{202}^c \{\chi_{0,1}^c H_{1,\frac{3}{2}}^c,\chi_{2,1}^c H_{1,\frac{3}{2}}^c,\chi_{2,2}^c H_{1,\frac{3}{2}}^c,\chi_{2,3}^c H_{1,\frac{3}{2}}^c\}$,~\\
$R_{022}^c \{\chi_{0,1}^c H_{1,\frac{3}{2}}^c,\chi_{2,1}^c H_{1,\frac{3}{2}}^c,\chi_{2,2}^c H_{1,\frac{3}{2}}^c,\chi_{2,3}^c H_{1,\frac{3}{2}}^c\}$\\
}}\\
\bottomrule[1pt]
\bottomrule[1pt]
\end{tabular}
\end{table*}

With the effective potentials of Eq.~(\ref{potential}), we could solve the three-body Schr\"odinger equation with the Gaussian expansion method. We not only calculate the binding energies, but also obtain the root-mean-square radii of $D^*$-$D^*$ and $D$-$D^*$. In general, orbitally excited hadronic molecular states are more difficult to be formed because of the repulsive centrifugal potential of the discussed systems. It is more likely to find bound state solutions from the $S$-wave ($l=L=0$) configurations in most situations. In the first step, we only consider the $S$-wave contributions. Then the $S$-$D$ mixing effect is included in the realistic calculation. When the $S$-$D$ mixing effect is  introduced, the tensor terms from the $\pi$, $\rho$, and $\omega$ contribute to the potential matrix elements. In the nuclear system, the tensor terms play an important role in the nucleon-nucleon interactions~\cite{Brown:1975di,Green:1974kh,Haapakoski:1974dr,Green:1975zm}. Similar results could be found in the charmed baryon-charmed baryon system~\cite{Meguro:2011nr}, where the tensor force from the $S$-$D$ mixing is necessary for obtaining the bound state solutions. Thus, in this work, we also consider the tensor terms. Besides the $S$-$D$ mixing and tensor terms, the coupled channel effect cannot be ignored when calculating the binding energy of a bound state~\cite{Li:2012ss,Tornqvist:1991ks}. Since both the $D^*D^*D$ and $D^*D^*D^*$ systems can have the quantum numbers $I(J^P)=\frac{1}{2}(0^-,1^-,2^-)$ and $I(J^P)=\frac{3}{2}(0^-,1^-,2^-)$, we should consider the $D^*D^*D$-$D^*D^*D^*$ coupled channel effect, which plays a role in the $DD^*\to D^*D^*$ or the $D^*D^*\to DD^*$ process.

In our approach, the cutoff $\Lambda$ is a crucial parameter when  searching for the bound state solutions of these discussed systems. With the measured binding energy of the $T_{cc}^+$ state, we obtained $\Lambda=0.976$, $0.998$, and $1.013$ GeV in Ref.~\cite{Wu:2021kbu}, which are close to the suggested values in previous works~\cite{Li:2012ss,Chen:2015loa,Chen:2019asm,Chen:2018pzd,Chen:2017jjn,Yasui:2009bz}. The $D^*D^*D$ and $D^*D^*D^*$ systems can be related to the $DDD^*$ system via heavy quark spin symmetry. As a result, in this work we will take the same strategy as that of Ref.~{\cite{Wu:2021kbu}}, i.e., we scan the range of $\Lambda$ from $0.90$ to $3.0$ GeV to search for bound states of $D^*D^*D$ and $D^*D^*D^*$. If a system has a bound state solution with $\Lambda\approx 1$ GeV, we view this state as a good molecular candidate.

In the study of hadronic molecular candidates composed of two charmed mesons~\cite{Li:2012ss}, it was found that systems with lower isospins are more likely to bind. In this work, we find that this is also true for systems composed of three charmed mesons.

\subsection{ $D^*D^*D$ system}

\begin{table*}[htbp]
\centering
\renewcommand\arraystretch{1.25}
\caption{Binding energies, root-mean-square radii, and probabilities of the $D^*D^*D$ system.}
\label{ResultDSDSDRes}
\begin{tabular}{p{0.4cm}<{\centering}p{0.4cm}<{\centering}
@{\vrule width 0.75pt}p{1.2cm}<{\centering}p{1.2cm}<{\centering}p{1.35cm}<{\centering}p{1.35cm}<{\centering}
@{\vrule width 0.75pt}p{1.2cm}<{\centering}p{1.2cm}<{\centering}p{1.35cm}<{\centering}p{1.35cm}<{\centering}
@{\vrule width 0.75pt}p{1.2cm}<{\centering}p{1.2cm}<{\centering}p{1.35cm}<{\centering}p{1.35cm}<{\centering}}
\toprule[1.00pt]
\toprule[1.00pt]
\multirow{2}{*}{$I$}&\multirow{2}{*}{$J^P$}&\multicolumn{4}{c@{\vrule width 0.75pt}}{$S$-wave}&\multicolumn{4}{c@{\vrule width 0.75pt}}{$S$-$D$ mixing}&\multicolumn{4}{c}{coupled channels}\\
\Xcline{3-14}{0.75pt}
&
&$\Lambda$ (GeV) &$B$ (MeV) &$r_{D^*D^*}$ (fm) &$r_{D^*D}$ (fm)
&$\Lambda$ (GeV) &$B$ (MeV) &$r_{D^*D^*}$ (fm) &$r_{D^*D}$ (fm)
&$\Lambda$ (GeV) &$B$ (MeV) &$P_{D^*D^*D}$(\%) &$P_{D^*D^*D^*}$(\%)\\
\midrule[0.75pt]
\multirow{9}{*}{$\frac{1}{2}$}
&\multirow{3}{*}{$0^-$}
 &0.98&0.45&7.73&6.11&0.95&0.44&7.64&6.08&0.92&0.84&99.34&0.66\\
&&1.03&3.60&3.91&3.03&1.00&3.38&3.81&2.98&0.97&5.44&97.95&2.05\\
&&1.08&9.65&2.55&1.99&1.05&9.10&2.51&1.97&1.02&14.79&95.86&4.14\\
\Xcline{2-14}{0.75pt}
&\multirow{3}{*}{$1^-$}
 &0.98&0.86&3.52&3.88&0.95&0.44&5.11&5.25&0.93&0.81&99.29&0.71\\
&&1.03&6.84&1.55&1.82&1.00&5.16&1.81&2.11&0.98&6.22&97.95&2.05\\
&&1.08&17.43&1.10&1.31&1.05&14.21&1.23&1.46&1.03&18.03&95.50&4.50\\
\Xcline{2-14}{0.75pt}
&\multirow{3}{*}{$2^-$}
 &0.97&0.82&4.36&3.67&0.94&0.57&5.34&4.45&0.90&0.48&99.26&0.74\\
&&1.02&6.30&2.00&1.71&0.99&5.04&2.28&1.94&0.95&7.05&96.32&3.68\\
&&1.07&16.19&1.41&1.22&1.04&13.54&1.56&1.35&1.00&22.16&91.42&8.58\\
\midrule[0.75pt]
\multirow{9}{*}{$\frac{3}{2}$}
&\multirow{3}{*}{$0^-$}
 &1.76&0.41&5.26&4.44&1.75&0.37&5.33&4.52&1.76&0.63&99.99&0.01\\
&&1.81&1.94&3.27&2.70&1.80&1.81&3.31&2.74&1.81&2.28&99.98&0.02\\
&&1.86&4.48&2.50&2.03&1.85&4.21&2.52&2.06&1.86&4.91&99.98&0.02\\
\Xcline{2-14}{0.75pt}
&\multirow{3}{*}{$1^-$}
 &1.85&0.46&11.75&8.68&1.85&0.54&11.98&8.80&1.85&0.57&$\sim$100&$\sim$0\\
&&1.90&2.20&11.23&8.09&1.90&2.41&1.88&14.38&1.90&2.41&$\sim$100&$\sim$0\\
&&1.95&4.97&10.93&7.81&1.95&5.61&1.35&14.26&1.95&5.61&$\sim$100&$\sim$0\\
\Xcline{2-14}{0.75pt}
&\multirow{3}{*}{$2^-$}
 &1.49&0.86&2.26&2.29&1.48&0.66&2.48&2.51&1.47&0.31&99.97&0.03\\
&&1.54&5.16&1.32&1.34&1.53&4.48&1.39&1.42&1.52&3.68&99.96&0.04\\
&&1.59&12.46&0.98&1.00&1.58&11.10&1.03&1.04&1.57&9.70&99.96&0.04\\
\bottomrule[1pt]
\bottomrule[1pt]
\end{tabular}
\end{table*}

For the $S$-wave only $D^*D^*D$ system, allowed spin parities are $0^-$, $1^-$, and $2^-$ with $I=\frac{1}{2}$ and $\frac{3}{2}$. In the $S$-$D$ mixing scheme, we require $l+L\leq 2$ to restrict the maximum orbital angular momentum. It should be noted that for the $D^*$-$D^*$ pair, the sum of $t+s+l$ should be odd. In Table~\ref{bases}, we present the configurations of the $D^*D^*D$-$D^*D^*D^*$ system. For the $S$-wave only and the $S$-$D$ mixing schemes of the $D^*D^*D$ system, we calculate the binding energies and root-mean-square radii for $I(J^P)=\frac{1}{2}(0^-,1^-,2^-)$ and $I(J^P)=\frac{3}{2}(0^-,1^-,2^-)$. In the  coupled-channel case of $D^*D^*D$-$D^*D^*D^*$, we present the binding energies and probabilities of the $D^*D^*D$ and $D^*D^*D^*$. The numerical results are shown in Table~\ref{ResultDSDSDRes}.

For the $S$-wave only $D^*D^*D$ system with $I(J^P)=\frac{1}{2}(0^-)$, we can obtain  bound state solutions when the cutoff $\Lambda$ reaches  about 0.98 GeV. The binding energy is on the order of MeV and the root-mean-square radii are several fm. If we increase the cutoff $\Lambda$, the binding energy increases while the root-mean-square radii decrease. We note that the consideration of $S$-$D$ mixing and the coupled-channel ($D^*D^*D^*$) effect increases the binding energy, which is reflected by the fact that a slightly smaller cutoff is needed to obtain a  binding energy similar to the case for which only the $S$-wave interaction is taken into account. For the cutoff range studied, the probability of the $D^*D^*D^*$ configuration is small and at the order of a few percent. Since the binding energy and root-mean-square radii are reasonable from the perspective of hadronic molecules, this state could be viewed as a good hadronic molecular candidate.

For the $S$-wave only $D^*D^*D$ system with $I(J^P)=\frac{1}{2}(1^-)$, we can also obtain  bound state solutions with the cutoff $\Lambda=0.98$, $1.03$, and $1.08$ GeV. Further consideration of the $S$-$D$ mixing and $D^*D^*D^*$ coupled-channel effect does not change the overall picture. According to the calculated binding energy and root-mean-square radii, this state  could also be treated as an ideal  hadronic molecular candidate.

For the $S$-wave only $D^*D^*D$ system with $I(J^P)=\frac{1}{2}(2^-)$, one can also find weakly bound states for the same cutoff $\Lambda$ as that of $I(J^P)=\frac{1}{2}(0^-,1^-)$. Similar to the case of $P_c(4440)$ and $P_c(4457)$, for the same cutoff, the $0^-,1^-,2^-$ states have different binding energies at the order of several hundred keV. With increased experimental precision, it is likely that these states can  be distinguished from each other in future experiments. The contribution of the $S$-$D$ mixing and coupled-channel effect is also similar to the case of $I(J^P)=\frac{1}{2}(0^-,1^-)$.

We also study the $D^*D^*D$ system for $I(J^P)=\frac{3}{2}(0^-,1^-,2^-)$. There are no bound state solutions for a cutoff $\Lambda$ below 1.013 GeV. In order to obtain bound state solutions for the $S$-wave  only $D^*D^*D$ systems with $I(J^P)=\frac{3}{2}(0^-,1^-2^-)$ , we increase the cutoff $\Lambda$ to $1.76\sim1.86$ GeV, $1.85\sim1.95$ GeV, and $1.49\sim1.59$ GeV, respectively. Here, we increase the cutoff in steps of $0.05$ GeV when scanning the cutoff $\Lambda$. Similar results are also obtained when the $S$-$D$ mixing and coupled channel effect are taken into account. Considering that the needed cutoff $\Lambda$ is out of the range of $0.976\sim1.013$ GeV (the range determined in Ref.~\cite{Wu:2021kbu}), we are a bit reluctant to view the $I(J^P)=\frac{3}{2}(0^-,1^-,2^-)$ $D^*D^*D$ bound states as good hadronic molecular candidates.

We note an interesting scenario in the $S$-$D$ mixing scheme for the $I(J^P)=\frac{3}{2}(1^-)$ case. When the cutoff $\Lambda$ changes from 1.85 GeV to 1.90 GeV, $r_{D^*D^*}$ decreases from 11.98 fm to 1.88 fm, while  $r_{D^*D}$ increases from 8.80 fm to 14.38 fm. For the $S$-$D$ mixing $I(J^P)=\frac{3}{2}(1^-)$ $D^*D^*D$ state, there are more than one bound states. For convenience, we use $s_1$ to denote the bound state solution with $r_{D^*D^*}\approx 11$ fm and $r_{D^*D}\approx 8$ fm, and  $s_2$ to label the bound state solution with $r_{D^*D^*}\approx 2$ fm and $r_{D^*D}\approx 14$ fm. The dependence of the two solutions $s_1$ and $s_2$ on the cutoff $\Lambda$ is found to be different. However, in Table~\ref{ResultDSDSDRes}, we only show the bound state solutions  with the largest binding energy. With $\Lambda=1.85$ GeV, we found that $B_{s1}>B_{s2}$; thus, we show the bound state solution $s_1$. While for $\Lambda=1.90$ and $1.95$ GeV, we present the bound state solution $s_2$ since $B_{s1}<B_{s2}$.

If only  the $S$-wave interaction had been considered for the $I(J^P)=\frac{3}{2}(1^-)$ $D^*D^*D$ state,  only  the bound state solution $s_1$ would have been obtained. Thus, the $S$-$D$ mixing effect plays a significant role for this state.  Further studies of the $S$-$D$ mixing effect shows that the bound state solution $s_2$ is highly correlated to the configuration $R_{022}^1\chi_{2,2}^1 H_{1,\frac{3}{2}}^1$ (see Table~\ref{bases}). This could be diagnosed in   the following steps:
\begin{enumerate}
\item In  the $S$-$D$ mixing scheme without the $R_{022}^1\chi_{2,2}^1 H_{1,\frac{3}{2}}^1$ configuration, the bound state solution $s_1$ exists but not the $s_2$ solution.
\item In the $S$-wave only combing with the $R_{022}^1\chi_{2,2}^1 H_{1,\frac{3}{2}}^1$ configuration, both solutions $s_1$ and $s_2$ exist.
\item If only the configuration $R_{022}^1\chi_{2,2}^1 H_{1,\frac{3}{2}}^1$ is considered, only the solution $s_2$ exists.
\end{enumerate}
From the above analysis, we conclude that the $R_{022}^1\chi_{2,2}^1 H_{1,\frac{3}{2}}^1$ configuration affects the results and contributes dominantly to the bound state solution $s_2$.

How to search for $D^*D^*D$ molecular candidates is also an interesting question. One possible decay mode is that the triple-charm molecules decay into a double-charm molecular state and a charmed meson. The other possible mode is that they directly decay into multibody final states bypassing intermediate states. Here, we summarize these channels as follows.
\begin{itemize}
\item If the binding energies are extremely small, they could first decay into $T_{cc}^+D^*$ , and then $T_{cc}^+$ can decay into $DD\pi$, and $D^*$ could be seen in the $D\pi$ and $D\gamma$ channels. In this case, the molecular candidates may be observed in the $DDD\pi\pi$ and $DDD\pi\gamma$ final states.
\item If the masses of the molecular candidates are below the $T_{cc}^+D^*$ threshold, the kinematically allowed channel is $T_{cc}^+D$. The $D^*D^*D$ molecular candidates could be studied in the $DDD\pi$ channel.
\item If the $D^*D^*$ hadronic molecular state exists, the $D^*D^*D$ molecular candidates may decay into a $D^*D^*$ molecular state and $D$. The $D^*D^*$ molecular state could be observed in the $D^*D$, $DD\pi$, and $DD\gamma$ final states.
\item In the above three scenarios, the $D^*D^*D$ molecular candidates ultimately decay into three charmed mesons together with $\pi$ and $\gamma$. We should also emphasize that these final states can originate not only from the intermediate double-charm molecular states with $D^{(*)}$, but also from nonresonant processes.
\item In addition, the $D^*D^*D$ molecular candidates can also decay into three charmed mesons via fall apart or quark rearrangement mechanisms. The typical channels are $D^*DD$ and $DDD$.
\end{itemize}
According to the discussions above, the $D^*D^*D$ molecular candidates could be studied with the three-, four-, and five-body final states in future experiments.

\subsection{ $D^*D^*D^*$ system}

\begin{table}[htbp]
\centering
\renewcommand\arraystretch{1.20}
\caption{Binding energies and root-mean-square radii of $D^*D^*D^*$ system.}
\label{ResultDSDSDSRes}
\begin{tabular}{@{\extracolsep{\fill}}cc
@{\vrule width 0.75pt}ccc
@{\vrule width 0.75pt}ccc}
\toprule[1.00pt]
\toprule[1.00pt]
\multirow{2}{*}{$I$}&\multirow{2}{*}{$J^P$}&\multicolumn{3}{c@{\vrule width 0.75pt}}{$S$-wave}&\multicolumn{3}{c}{$S$-$D$ mixing}\\
\Xcline{3-8}{0.75pt}
&
&$\Lambda$ (GeV) &$B$ (MeV) &$r_{D^*D^*}$ (fm)
&$\Lambda$ (GeV) &$B$ (MeV) &$r_{D^*D^*}$ (fm)\\
\midrule[0.75pt]
\multirow{12}{*}{$\frac{1}{2}$}
&\multirow{3}{*}{$0^-$}
 &1.02&0.67&9.75&1.00&0.84&9.92\\
&&1.07&4.48&9.06&1.05&4.57&9.17\\
&&1.12&10.72&8.81&1.10&10.63&8.88\\
\Xcline{2-8}{0.75pt}
&\multirow{3}{*}{$1^-$}
 &1.00&0.67&5.52&0.97&0.44&6.02\\
&&1.05&5.02&2.51&1.02&4.30&2.61\\
&&1.10&12.83&1.73&1.07&11.52&1.78\\
\Xcline{2-8}{0.75pt}
&\multirow{3}{*}{$2^-$}
 &0.98&0.37&4.71&0.96&0.51&4.53\\
&&1.03&5.94&1.79&1.01&5.80&1.87\\
&&1.08&16.33&1.25&1.06&15.55&1.31\\
\Xcline{2-8}{0.75pt}
&\multirow{3}{*}{$3^-$}
 &1.84&0.51&9.74&1.02&0.35&12.19\\
&&1.89&2.39&9.18&1.07&4.03&11.72\\
&&1.94&5.38&8.90&1.12&10.23&11.63\\
\midrule[0.75pt]
\multirow{9}{*}{$\frac{3}{2}$}
&\multirow{3}{*}{$1^-$}
 &1.80&0.20&7.82&1.81&0.46&7.21\\
&&1.85&1.40&6.11&1.86&1.89&5.59\\
&&1.90&3.59&4.92&1.91&4.37&4.45\\
\Xcline{2-8}{0.75pt}
&\multirow{3}{*}{$2^-$}
 &1.85&0.81&9.60&1.84&0.61&9.89\\
&&1.90&2.89&9.13&1.89&2.51&9.29\\
&&1.95&6.13&8.88&1.94&5.53&8.98\\
\Xcline{2-8}{0.75pt}
&\multirow{3}{*}{$3^-$}
 &1.48&0.23&2.95&1.48&0.78&2.31\\
&&1.53&4.21&1.38&1.53&4.90&1.34\\
&&1.58&11.28&1.00&1.58&11.93&1.00\\
\bottomrule[1pt]
\bottomrule[1pt]
\end{tabular}
\end{table}

Since the $D^*D^*D^*$ system contains three identical mesons, the $c=1,2,3$ channels share the same configurations. In addition, for all the channels, $(-1)^{t+s+l+1}=1$, which restricts allowed combinations of $t$, $s$, and $l$. For the $S$-wave only $D^*D^*D^*$ system with $I=\frac{1}{2}$, the allowed spin parities are $0^-$, $1^-$, $2^-$, and $3^-$. For  $I=\frac{3}{2}$ in $S$-wave, the allowed spin parities are $1^-$, $2^-$, and $3^-$. For all the $D^*D^*D^*$ states, we also consider the $S$-$D$ mixing effect. As shown in Table~\ref{ResultDSDSDRes}, the coupled-channel effects between $D^*D^*D$-$D^*D^*D^*$ are  small and thus could be neglected. We notice that the threshold of $D^*D^*D^*$ is about 140 MeV higher than that of  $D^*D^*D$, and therefore the $D^*D^*D$ component is difficult to be bounded in the $D^*D^*D^*$-predominate states. In general, the coupled-channel effects of $D^*D^*D$-$D^*D^*D^*$ mainly affect the decay behaviors of the $D^*D^*D^*$ states. Since we focus primarily on the existences of the bound states of  $D^*D^*D^*$, the coupled-channel effects of $D^*D^*D$-$D^*D^*D^*$ are not considered here. The binding energies and root-mean-square radii are  presented in Table~\ref{ResultDSDSDSRes}.

For the $S$-wave only $D^*D^*D^*$ system with $I(J^P)=\frac{1}{2}(0^-)$, we find a bound state solution for a cutoff $\Lambda$ larger than 1.02 GeV. The root-mean-square radii decrease slowly with the increase of the cutoff $\Lambda$. The radius $r_{D^*D^*}$ is estimated to be about 9 fm with $\Lambda\sim1$ GeV, which is a bit larger than that of $T_{cc}^+$ but similar to those of the $DDD^*$ states. Judging from the binding energy and root-mean-square radii, this state could be viewed as a good hadronic molecular candidate.

For the $S$-wave only $D^*D^*D^*$ system with $I(J^P)=\frac{1}{2}(1^-)$, we obtain a binding energy in the range of $0.67\sim12.83$ MeV for a cutoff $\Lambda$ between 1.00 and 1.10 GeV. While the root-mean-square radii decrease from 5.52 to 1.73 fm with the increase of the cutoff $\Lambda$.

For the $S$-wave only $D^*D^*D^*$ system with $I(J^P)=\frac{1}{2}(2^-)$, the system becomes bound when the cutoff $\Lambda$ reaches about 0.98 GeV. Since the obtained binding energy is on the order of MeV and the root-mean-square radii are several fm, this state is also an ideal hadronic molecular candidate.

For the three configurations studied, turning on the $S$-$D$ mixing  only has a small effect but, in general, slightly increases the binding energy of the system of interest (for the same cutoff).

For the $S$-wave only $D^*D^*D^*$ system with $I(J^P)=\frac{1}{2}(3^-)$, there are no bound state solution with a $\Lambda\approx1$ GeV. But if  the $S$-$D$ mixing is taken into account, we can obtain loosely bound state solutions for a cutoff $\Lambda\approx1$ GeV. By carefully studying the configurations in the $S$-$D$ mixing scheme for the case of $I(J^P)=\frac{1}{2}(3^-)$ $D^*D^*D^*$, we find that the configurations of $R_{022}^c \chi_{1,1}^c H_{0,\frac{1}{2}}^c$ and $R_{022}^c \chi_{1,2}^c H_{0,\frac{1}{2}}^c$ ($c=1,2,3$) play a key role in forming bound states. Similar to the analysis performed in studying the $I(J^P)=\frac{3}{2}(2^-)$ $D^*D^*D$ state, the above conclusion is obtained in the following way
\begin{enumerate}
\item In  the $S$-$D$ mixing scheme without $R_{022}^c \chi_{1,1}^c H_{0,\frac{1}{2}}^c$ and $R_{022}^c \chi_{1,2}^c H_{0,\frac{1}{2}}^c$ ($c=1,2,3$), there is no bound state solutions with $\Lambda\approx1$ GeV.
\item In the $S$-wave only combing with $R_{022}^c \chi_{1,1}^c H_{0,\frac{1}{2}}^c$ and $R_{022}^c \chi_{1,2}^c H_{0,\frac{1}{2}}^c$ ($c=1,2,3$), it is easy to find bound state solutions with $\Lambda\approx1$ GeV and the binding energies are approximate to those in  the $S$-$D$ mixing scheme given in Table~\ref{ResultDSDSDSRes}.
\item If the $R_{022}^c \chi_{1,1}^c H_{0,\frac{1}{2}}^c$ or $R_{022}^c \chi_{1,2}^c H_{0,\frac{1}{2}}^c$ ($c=1,2,3$) configuration is considered, nearly the same binding energy is obtained as that of the $S$-$D$ mixing scheme in Table~\ref{ResultDSDSDSRes} when $\Lambda\approx1$ GeV.
\end{enumerate}

Because of the complexity of the three-body problem, it is difficult to present a precise interpretation for this phenomenon, but some qualitative analyses are helpful to understand the  numerical results. For the configurations $R_{022}^c \chi_{1,1}^c H_{0,\frac{1}{2}}^c$ and $R_{022}^c \chi_{1,2}^c H_{0,\frac{1}{2}}^c$ ($c=1,2,3$), the isospin and spin in the ${\bf r}_c$ degree of freedom are $t=0$ and $s=1$, respectively, and the flavor and spin factors of the $\pi$ exchange are ${\cal C}_1(0)=-3/2$ and $\langle {\cal O}_7\rangle=1$~\cite{Liu:2019stu}, respectively. In this spin-isospin configuration, the $D^*$-$D^*$ force from the $\pi$ exchange is attractive and about 3 times of that in the $S$-wave only scheme with $t=1$ and $s=2$. We  notice that  $R_{022}^1\chi_{2,2}^1 H_{1,\frac{3}{2}}^1$ in the $D^*D^*D$ system, and $R_{022}^c \chi_{1,1}^c H_{0,\frac{1}{2}}^c$ and $R_{022}^c \chi_{1,2}^c H_{0,\frac{1}{2}}^c$ ($c=1,2,3$) in the $D^*D^*D^*$ system are ${\bf R}_c$-mode excited configurations, i.e., $l=0$, $L=2$ ($l=2$, $L=0$ for the ${\bf r}_c$-mode $D$-wave excited configuration.) Since the reduced mass of the ${\bf R}_c$ degree of freedom is larger than that of the ${\bf r}_c$ degree of freedom, if we take the same Gaussian variational parameters as inputs, the ${\bf R}_c$-mode excited configuration has a smaller kinetic matrix element than that of the ${\bf r}_c$-mode excited configuration, which is beneficial to form a ${\bf R}_c$-mode excited state. This might be the reason why some ${\bf R}_c$-mode excited configurations in the $D^*D^*D$ and $D^*D^*D^*$ systems play a significant role in the $S$-$D$ mixing scheme.

The three-body system contains two spatial degree of freedoms. If we introduce $S$-$D$ mixing, a large number of configurations are included in the calculation. For some specific spin-isospin configurations,  the $\pi$ exchange force is attractive. If such spin-flavor configurations emerge  in the $S$-$D$ mixing scheme but not  in the $S$-wave only scenario, we should carefully investigate the $S$-$D$ mixing effect.

According to the above analysis, it is possible to find $D$-wave bound state solutions. But in the present work, we  mainly focus on molecular states in the $S$-wave only or $S$-$D$ mixing scenarios.

For the $S$-wave only $I(J^P)=\frac{3}{2}(1^-,2^-,3^-)$ $D^*D^*D^*$ systems, a larger cutoff is needed for them to bind. More specifically, we only find  bound state solutions when the cutoff $\Lambda$ reaches around 1.8 GeV for $I(J^P)=\frac{3}{2}(1^-,2^-)$. We also obtain shallow bound states in the $S$-wave only $I(J^P)=\frac{3}{2}(3^-)$ $D^*D^*D^*$ system if the cutoff $\Lambda$ is close to 1.5 GeV. Since the needed cutoff $\Lambda$ is much larger than our expectation, we prefer not to view these states as good hadronic molecular candidates. For all the $I=3/2$ configurations, the $S$-$D$ mixing effect is relatively small and plays a minor role.

Since the $D^*D^*D^*$ molecular candidates have larger masses than that of the $D^*D^*D$ system, much more complex decay modes can be anticipated. Here, the decay channels are summarized as the following:
\begin{itemize}
\item In principle, all the decay modes of the $D^*D^*D$ molecular candidates are also kinematically allowed for the $D^*D^*D^*$ system.
\item There are also some modes  specific for the $D^*D^*D^*$ system. For example, the channel of a $D^*D^*$ molecular candidate with a $D^*$ meson is only kinematically allowed for the $D^*D^*D^*$ system.
\end{itemize}
However, although there are more decay channels for the $D^*D^*D^*$ system, the decay patterns are similar for the $DDD^*$~\cite{Wu:2021kbu}, $D^*D^*D$, and $D^*D^*D^*$ molecular states. In the future, these states could be searched for by measuring three charmed mesons together with pions and photons in the final states.

\begin{table*}
\caption{Dependence of binding energies on the variation  of the coupling constants by about 10\% in the $D^*D^*D$ system. The cutoff $\Lambda$, potential expectations $\langle V\rangle$, and binding energies $B^{(\prime)}$ are in units of GeV, MeV, and MeV, respectively. The binding energies $B$ are calculated with the values of the coupling constants given in the fifth column, and the $B^\prime$ are obtained within their uncertainties.}
\label{undetermineddsdsd}
\centering
\renewcommand\arraystretch{1.25}
\scalebox{0.8}{\begin{tabular}{ccccc@{\vrule width 0.75pt}ccccc@{\vrule width 0.75pt}ccccc@{\vrule width 0.75pt}ccccc}
\toprule[1.00pt]
\toprule[1.00pt]
 \multirow{2}[5]{*}{$I$}
&\multirow{2}[5]{*}{$J^P$}
&\multirow{2}[5]{*}{\makecell[c]{exchanged\\mesons}}
&\multirow{2}[5]{*}{\makecell[c]{coupling\\constant}}
&\multirow{2}[5]{*}{\makecell[c]{reference\\value}}
&\multicolumn{5}{c@{\vrule width 0.75pt}}{$S$-wave}
&\multicolumn{5}{c@{\vrule width 0.75pt}}{$S$-$D$ mixing}
&\multicolumn{5}{c}{Coupled channel}\\
\Xcline{6-20}{0.75pt}
&&&&&$\Lambda$&$B$&$\langle V\rangle$&\makecell[c]{reference\\range}&$B^\prime$
    &$\Lambda$&$B$&$\langle V\rangle$&\makecell[c]{reference\\range}&$B^\prime$
    &$\Lambda$&$B$&$\langle V\rangle$&\makecell[c]{reference\\range}&$B^\prime$\\
\midrule[0.75pt]
\multirow{15}{*}{$\frac{1}{2}$}
&\multirow{5}{*}{$0^-$}
 &$\pi$         &$g$                       &0.6  &\multirow{5}{*}{1.03}&\multirow{5}{*}{3.60}&-16.81&0.54$\sim$0.66&1.08$\sim$7.92&\multirow{5}{*}{1.00}&\multirow{5}{*}{3.38}&-19.16&0.54$\sim$0.66&0.68$\sim$8.51&\multirow{5}{*}{0.97}&\multirow{5}{*}{5.44}&-32.83&0.54$\sim$0.66&0.99$\sim$14.58\\
&&$\sigma$      &$g_\sigma$                &3.4  &                     &                      &-15.96&3.06$\sim$3.74&1.27$\sim$7.87&                     &                      &-13.73&3.06$\sim$3.74&1.34$\sim$6.99&                     &                      &-14.95&3.06$\sim$3.74&3.02$\sim$9.10\\
&&$\rho,~\omega$&$\beta g_V$               &5.2  &                     &                      &-3.98&4.68$\sim$5.72&2.87$\sim$4.47&                     &                      &-2.93&4.68$\sim$5.72&2.84$\sim$4.02&                     &                      &-3.15&4.68$\sim$5.72&4.86$\sim$6.12\\
&&$\rho,~\omega$&$\lambda g_V$~(GeV$^{-1}$)&3.133&                     &                      &-1.63&2.82$\sim$3.45&3.30$\sim$3.95&                     &                      &-0.84&2.82$\sim$3.45&3.22$\sim$3.56&                     &                      &-2.00&2.82$\sim$3.45&5.08$\sim$5.88\\
&&\multicolumn{3}{c@{\vrule width 0.75pt}}{kinetic energy $\langle T\rangle$ (MeV)}&&&34.78&$\cdots$&$\cdots$&&&33.28&$\cdots$&$\cdots$&&&47.48&$\cdots$&$\cdots$\\
\Xcline{2-20}{0.75pt}
&\multirow{5}{*}{$1^-$}
 &$\pi$         &$g$                       &0.6  &\multirow{5}{*}{1.03}&\multirow{5}{*}{6.84}&-30.71&0.54$\sim$0.66&2.17$\sim$14.68&\multirow{5}{*}{1.00}&\multirow{5}{*}{5.16}&-28.87&0.54$\sim$0.66&1.01$\sim$12.76&\multirow{5}{*}{0.98}&\multirow{5}{*}{6.22}&-38.64&0.54$\sim$0.66&1.07$\sim$17.08\\
&&$\sigma$      &$g_\sigma$                &3.4  &                     &                      &-32.49&3.06$\sim$3.74&1.93$\sim$14.95&                     &                      &-24.40&3.06$\sim$3.74&1.52$\sim$11.34&                     &                      &-22.92&3.06$\sim$3.74&3.17$\sim$12.16\\
&&$\rho,~\omega$&$\beta g_V$               &5.2  &                     &                      &-3.83&4.68$\sim$5.72&6.14$\sim$7.68&                     &                      &-2.82&4.68$\sim$5.72&4.64$\sim$5.77&                     &                      &-3.08&4.68$\sim$5.72&5.66$\sim$6.89\\
&&$\rho,~\omega$&$\lambda g_V$~(GeV$^{-1}$)&3.133&                     &                      &-3.83&2.82$\sim$3.45&6.14$\sim$7.68&                     &                      &-1.98&2.82$\sim$3.45&4.79$\sim$5.59&                     &                      &-2.77&2.82$\sim$3.45&5.72$\sim$6.83\\
&&\multicolumn{3}{c@{\vrule width 0.75pt}}{kinetic energy $\langle T\rangle$ (MeV)}&&&64.02&$\cdots$&$\cdots$&&&52.91&$\cdots$&$\cdots$&&&61.19&$\cdots$&$\cdots$\\
\Xcline{2-20}{0.75pt}
&\multirow{5}{*}{$2^-$}
 &$\pi$         &$g$                       &0.6  &\multirow{5}{*}{1.02}&\multirow{5}{*}{6.30}&-29.65&0.54$\sim$0.66&1.84$\sim$13.91&\multirow{5}{*}{0.99}&\multirow{5}{*}{5.04}&-28.61&0.54$\sim$0.66&0.95$\sim$12.58&\multirow{5}{*}{0.95}&\multirow{5}{*}{7.05}&-50.13&0.54$\sim$0.66&0.70$\sim$21.63\\
&&$\sigma$      &$g_\sigma$                &3.4  &                     &                      &-29.67&3.06$\sim$3.74&1.84$\sim$13.76&                     &                      &-22.48&3.06$\sim$3.74&1.66$\sim$10.74&                     &                      &-25.05&3.06$\sim$3.74&3.05$\sim$13.14\\
&&$\rho,~\omega$&$\beta g_V$               &5.2  &                     &                      &-3.34&4.68$\sim$5.72&5.69$\sim$7.03&                     &                      &-2.43&4.68$\sim$5.72&4.59$\sim$5.56&                     &                      &-2.28&4.68$\sim$5.72&6.63$\sim$7.54\\
&&$\rho,~\omega$&$\lambda g_V$~(GeV$^{-1}$)&3.133&                     &                      &-3.30&2.82$\sim$3.45&5.70$\sim$7.03&                     &                      &-1.67&2.82$\sim$3.45&4.73$\sim$5.40&                     &                      &-3.10&2.82$\sim$3.45&6.49$\sim$7.73\\
&&\multicolumn{3}{c@{\vrule width 0.75pt}}{kinetic energy $\langle T\rangle$ (MeV)}&&&59.65&$\cdots$&$\cdots$&&&50.15&$\cdots$&$\cdots$&&&73.51&$\cdots$&$\cdots$\\
\midrule[0.75pt]
\multirow{15}{*}{$\frac{3}{2}$}
&\multirow{5}{*}{$0^-$}
 &$\pi$         &$g$                       &0.6  &\multirow{5}{*}{1.90}&\multirow{5}{*}{7.31}&-32.13&0.54$\sim$0.66&2.31$\sim$15.32&\multirow{5}{*}{1.89}&\multirow{5}{*}{6.91}&-31.98&0.54$\sim$0.66&2.09$\sim$15.20&\multirow{5}{*}{1.89}&\multirow{5}{*}{7.01}&-32.18&0.54$\sim$0.66&2.21$\sim$15.49\\
&&$\sigma$      &$g_\sigma$                &3.4  &                     &                      &-92.22&3.20$\sim$3.74&0.15$\sim$35.37&                     &                      &-88.94&3.20$\sim$3.74&0.07$\sim$34.14&                     &                      &-89.34&3.20$\sim$3.74&0.10$\sim$34.31\\
&&$\rho,~\omega$&$\beta g_V$               &5.2  &                     &                      &69.22&4.68$\sim$5.51&24.84$\sim$0.96&                     &                      &66.51&4.68$\sim$5.51&23.82$\sim$0.84&                     &                      &66.81&4.68$\sim$5.51&23.97$\sim$0.89\\
&&$\rho,~\omega$&$\lambda g_V$~(GeV$^{-1}$)&3.133&                     &                      &-54.62&2.82$\sim$3.45&0.51$\sim$23.34&                     &                      &-51.05&2.82$\sim$3.45&0.58$\sim$22.25&                     &                      &-51.41&2.82$\sim$3.45&0.67$\sim$22.65\\
&&\multicolumn{3}{c@{\vrule width 0.75pt}}{kinetic energy $\langle T\rangle$ (MeV)}&&&102.43&$\cdots$&$\cdots$&&&98.56&$\cdots$&$\cdots$&&&99.12&$\cdots$&$\cdots$\\
\Xcline{2-20}{0.75pt}
&\multirow{5}{*}{$1^-$}
 &$\pi$         &$g$                       &0.6  &\multirow{5}{*}{1.98}&\multirow{5}{*}{7.23}&-34.29&0.54$\sim$0.66&2.05$\sim$16.04&\multirow{5}{*}{1.98}&\multirow{5}{*}{8.19}&-37.21&0.54$\sim$0.66&2.49$\sim$17.62&\multirow{5}{*}{1.98}&\multirow{5}{*}{8.19}&-37.21&0.54$\sim$0.66&2.49$\sim$17.62\\
&&$\sigma$      &$g_\sigma$                &3.4  &                     &                      &-73.64&3.20$\sim$3.74&0.66$\sim$27.31&                     &                      &-81.33&3.20$\sim$3.74&0.79$\sim$29.84&                     &                      &-81.33&3.20$\sim$3.74&0.80$\sim$29.84\\
&&$\rho,~\omega$&$\beta g_V$               &5.2  &                     &                      &58.45&4.68$\sim$5.62&21.03$\sim$0.20&                     &                      &64.81&4.68$\sim$5.62&23.20$\sim$0.29&                     &                      &64.81&4.68$\sim$5.62&23.20$\sim$0.33\\
&&$\rho,~\omega$&$\lambda g_V$~(GeV$^{-1}$)&3.133&                     &                      &-60.69&2.82$\sim$3.45&0.09$\sim$26.15&                     &                      &-68.38&2.82$\sim$3.45&0.20$\sim$29.03&                     &                      &-68.38&2.82$\sim$3.45&0.25$\sim$29.03\\
&&\multicolumn{3}{c@{\vrule width 0.75pt}}{kinetic energy $\langle T\rangle$ (MeV)}&&&102.95&$\cdots$&$\cdots$&&&113.92&$\cdots$&$\cdots$&&&113.92&$\cdots$&$\cdots$\\
\Xcline{2-20}{0.75pt}
&\multirow{5}{*}{$2^-$}
 &$\pi$         &$g$                       &0.6  &\multirow{5}{*}{1.62}&\multirow{5}{*}{18.54}&-82.25&0.54$\sim$0.66&5.67$\sim$38.92&\multirow{5}{*}{1.63}&\multirow{5}{*}{21.18}&-89.63&0.54$\sim$0.66&7.14$\sim$43.46&\multirow{5}{*}{1.62}&\multirow{5}{*}{19.01}&-85.33&0.54$\sim$0.66&5.86$\sim$40.52\\
&&$\sigma$      &$g_\sigma$                &3.4  &                     &                      &-201.84&3.20$\sim$3.74&0.19$\sim$70.36&                     &                      &-212.12&3.20$\sim$3.74&1.32$\sim$74.98&                     &                      &-201.46&3.20$\sim$3.74&0.57$\sim$70.73\\
&&$\rho,~\omega$&$\beta g_V$               &5.2  &                     &                      &143.99&4.68$\sim$5.62&50.62$\sim$0.23&                     &                      &152.23&4.68$\sim$5.62&54.79$\sim$1.27&                     &                      &143.70&4.68$\sim$5.62&51.03$\sim$0.61\\
&&$\rho,~\omega$&$\lambda g_V$~(GeV$^{-1}$)&3.133&                     &                      &-108.11&2.82$\sim$3.45&3.62$\sim$48.58&                     &                      &-115.54&2.82$\sim$3.45&5.20$\sim$53.45&                     &                      &-105.39&2.82$\sim$3.45&4.61$\sim$48.87\\
&&\multicolumn{3}{c@{\vrule width 0.75pt}}{kinetic energy $\langle T\rangle$ (MeV)}&&&229.67&$\cdots$&$\cdots$&&&243.87&$\cdots$&$\cdots$&&&229.47&$\cdots$&$\cdots$\\
\bottomrule[1pt]
\bottomrule[1pt]
\end{tabular}}
\end{table*}

\begin{table*}
\caption{Dependence of binding energies  on the variation  of the coupling constants by about 10\% for the $D^*D^*D^*$ system. The cutoff $\Lambda$, potential expectations $\langle V\rangle$, and binding energies $B^{(\prime)}$ are in units of GeV, MeV, and MeV, respectively.  The binding energies $B$ are calculated with the values of the coupling constant values given in the fifth column, and the $B^\prime$ are obtained within their uncertainties.}
\label{undetermineddsdsds}
\centering
\renewcommand\arraystretch{1.25}
\begin{tabular}{ccccc@{\vrule width 0.75pt}ccccc@{\vrule width 0.75pt}ccccc}
\toprule[1.00pt]
\toprule[1.00pt]
 \multirow{2}[5]{*}{$I$}
&\multirow{2}[5]{*}{$J^P$}
&\multirow{2}[5]{*}{\makecell[c]{exchanged\\mesons}}
&\multirow{2}[5]{*}{\makecell[c]{coupling\\constant}}
&\multirow{2}[5]{*}{\makecell[c]{reference\\value}}
&\multicolumn{5}{c@{\vrule width 0.75pt}}{$S$-wave}
&\multicolumn{5}{c}{$S$-$D$ mixing}\\
\Xcline{6-15}{0.75pt}
&&&&&$\Lambda$&$B$&$\langle V\rangle$&\makecell[c]{reference\\range}&$B^\prime$
    &$\Lambda$&$B$&$\langle V\rangle$&\makecell[c]{reference\\range}&$B^\prime$\\
\midrule[0.75pt]
\multirow{20}{*}{$\frac{1}{2}$}
&\multirow{5}{*}{$0^-$}
 &$\pi$         &$g$                       &0.6  &\multirow{5}{*}{1.07}&\multirow{5}{*}{4.48}&-17.57&0.54$\sim$0.66&1.70$\sim$8.87&\multirow{5}{*}{1.05}&\multirow{5}{*}{4.57}&-20.01&0.54$\sim$0.66&1.47$\sim$9.64\\
&&$\sigma$      &$g_\sigma$                &3.4  &                     &                      &-15.00&3.06$\sim$3.74&1.94$\sim$7.97&                     &                      &-13.55&3.06$\sim$3.74&2.25$\sim$7.70\\
&&$\rho,~\omega$&$\beta g_V$               &5.2  &                     &                      &-6.49&4.68$\sim$5.72&3.30$\sim$5.91&                     &                      &-5.56&4.68$\sim$5.72&3.56$\sim$5.79\\
&&$\rho,~\omega$&$\lambda g_V$~(GeV$^{-1}$)&3.133&                     &                      &-2.40&2.82$\sim$3.45&4.03$\sim$5.00&                     &                      &-1.60&2.82$\sim$3.45&4.27$\sim$4.92\\
&&\multicolumn{3}{c@{\vrule width 0.75pt}}{kinetic energy $\langle T\rangle$ (MeV)}&&&36.99&$\cdots$&$\cdots$&&&36.16&$\cdots$&$\cdots$\\
\Xcline{2-15}{0.75pt}
&\multirow{5}{*}{$1^-$}
 &$\pi$         &$g$                       &0.6  &\multirow{5}{*}{1.05}&\multirow{5}{*}{5.02}&-20.15&0.54$\sim$0.66&1.88$\sim$10.07&\multirow{5}{*}{1.02}&\multirow{5}{*}{4.30}&-22.03&0.54$\sim$0.66&1.08$\sim$10.05\\
&&$\sigma$      &$g_\sigma$                &3.4  &                     &                      &-23.45&3.06$\sim$3.74&1.61$\sim$11.18&                     &                      &-19.51&3.06$\sim$3.74&1.44$\sim$9.40\\
&&$\rho,~\omega$&$\beta g_V$               &5.2  &                     &                      &-5.05&4.68$\sim$5.72&4.09$\sim$6.12&                     &                      &-3.73&4.68$\sim$5.72&3.62$\sim$5.12\\
&&$\rho,~\omega$&$\lambda g_V$~(GeV$^{-1}$)&3.133&                     &                      &-2.54&2.82$\sim$3.45&4.55$\sim$5.57&                     &                      &-1.35&2.82$\sim$3.45&4.05$\sim$4.59\\
&&\multicolumn{3}{c@{\vrule width 0.75pt}}{kinetic energy $\langle T\rangle$ (MeV)}&&&46.17&$\cdots$&$\cdots$&&&42.31&$\cdots$&$\cdots$\\
\Xcline{2-15}{0.75pt}
&\multirow{5}{*}{$2^-$}
 &$\pi$         &$g$                       &0.6  &\multirow{5}{*}{1.03}&\multirow{5}{*}{5.94}&-27.90&0.54$\sim$0.66&1.74$\sim$13.11&\multirow{5}{*}{1.01}&\multirow{5}{*}{5.80}&-30.15&0.54$\sim$0.66&1.36$\sim$13.62\\
&&$\sigma$      &$g_\sigma$                &3.4  &                     &                      &-31.72&3.06$\sim$3.74&1.29$\sim$13.98&                     &                      &-27.52&3.06$\sim$3.74&1.66$\sim$12.73\\
&&$\rho,~\omega$&$\beta g_V$               &5.2  &                     &                      &-3.71&4.68$\sim$5.72&5.27$\sim$6.76&                     &                      &-3.10&4.68$\sim$5.72&5.23$\sim$6.47\\
&&$\rho,~\omega$&$\lambda g_V$~(GeV$^{-1}$)&3.133&                     &                      &-3.66&2.82$\sim$3.45&5.28$\sim$6.75&                     &                      &-2.33&2.82$\sim$3.45&5.37$\sim$6.30\\
&&\multicolumn{3}{c@{\vrule width 0.75pt}}{kinetic energy $\langle T\rangle$ (MeV)}&&&61.04&$\cdots$&$\cdots$&&&57.30&$\cdots$&$\cdots$\\
\Xcline{2-15}{0.75pt}
&\multirow{5}{*}{$3^-$}
 &$\pi$         &$g$                       &0.6  &\multirow{5}{*}{1.96}&\multirow{5}{*}{6.95}&-32.88&0.54$\sim$0.66&1.98$\sim$15.40&\multirow{5}{*}{1.07}&\multirow{5}{*}{4.03}&-17.33&0.54$\sim$0.66&1.32$\sim$8.39\\
&&$\sigma$      &$g_\sigma$                &3.4  &                     &                      &-73.68&3.20$\sim$3.74&0.44$\sim$27.08&                     &                      &-14.81&3.06$\sim$3.74&1.55$\sim$7.49\\
&&$\rho,~\omega$&$\beta g_V$               &5.2  &                     &                      &58.27&4.68$\sim$5.62&20.73$\sim$0.02&                     &                      &-6.40&4.68$\sim$5.72&2.88$\sim$5.45\\
&&$\rho,~\omega$&$\lambda g_V$~(GeV$^{-1}$)&3.133&                     &                      &-59.30&2.88$\sim$3.45&0.71$\sim$25.39&                     &                      &-2.37&2.82$\sim$3.45&3.60$\sim$4.55\\
&&\multicolumn{3}{c@{\vrule width 0.75pt}}{kinetic energy $\langle T\rangle$ (MeV)}&&&100.65&$\cdots$&$\cdots$&&&36.88&$\cdots$&$\cdots$\\
\midrule[0.75pt]
\multirow{15}{*}{$\frac{3}{2}$}
&\multirow{5}{*}{$1^-$}
 &$\pi$         &$g$                       &0.6  &\multirow{5}{*}{1.93}&\multirow{5}{*}{5.48}&-27.36&0.54$\sim$0.66&1.46$\sim$12.64&\multirow{5}{*}{1.94}&\multirow{5}{*}{6.45}&-29.86&0.54$\sim$0.66&2.07$\sim$14.40\\
&&$\sigma$      &$g_\sigma$                &3.4  &                     &                      &-67.77&3.20$\sim$3.74&0.16$\sim$26.14&                     &                      &-73.22&3.20$\sim$3.74&0.46$\sim$28.15\\
&&$\rho,~\omega$&$\beta g_V$               &5.2  &                     &                      &52.33&4.68$\sim$5.51&18.73$\sim$0.72&                     &                      &56.68&4.68$\sim$5.62&20.49$\sim$0.19\\
&&$\rho,~\omega$&$\lambda g_V$~(GeV$^{-1}$)&3.133&                     &                      &-48.96&2.82$\sim$3.45&0.09$\sim$21.14&                     &                      &-54.19&2.82$\sim$3.45&0.35$\sim$23.87\\
&&\multicolumn{3}{c@{\vrule width 0.75pt}}{kinetic energy $\langle T\rangle$ (MeV)}&&&86.28&$\cdots$&$\cdots$&&&94.14&$\cdots$&$\cdots$\\
\Xcline{2-15}{0.75pt}
&\multirow{5}{*}{$2^-$}
 &$\pi$         &$g$                       &0.6  &\multirow{5}{*}{1.97}&\multirow{5}{*}{7.81}&-35.10&0.54$\sim$0.66&2.45$\sim$16.75&\multirow{5}{*}{1.96}&\multirow{5}{*}{7.11}&-32.58&0.54$\sim$0.66&2.25$\sim$15.67\\
&&$\sigma$      &$g_\sigma$                &3.4  &                     &                      &-77.67&3.20$\sim$3.74&0.80$\sim$28.71&                     &                      &-73.88&3.20$\sim$3.74&0.55$\sim$27.26\\
&&$\rho,~\omega$&$\beta g_V$               &5.2  &                     &                      &61.67&4.68$\sim$5.62&22.21$\sim$0.30&                     &                      &58.44&4.68$\sim$5.62&20.91$\sim$0.13\\
&&$\rho,~\omega$&$\lambda g_V$~(GeV$^{-1}$)&3.133&                     &                      &-63.91&2.82$\sim$3.45&0.19$\sim$27.45&                     &                      &-60.23&2.82$\sim$3.45&0.14$\sim$26.29\\
&&\multicolumn{3}{c@{\vrule width 0.75pt}}{kinetic energy $\langle T\rangle$ (MeV)}&&&107.21&$\cdots$&$\cdots$&&&101.14&$\cdots$&$\cdots$\\
\Xcline{2-15}{0.75pt}
&\multirow{5}{*}{$3^-$}
 &$\pi$         &$g$                       &0.6  &\multirow{5}{*}{1.62}&\multirow{5}{*}{19.55}&-83.42&0.54$\sim$0.66&6.40$\sim$40.13&\multirow{5}{*}{1.62}&\multirow{5}{*}{20.12}&-87.08&0.54$\sim$0.66&6.62$\sim$41.99\\
&&$\sigma$      &$g_\sigma$                &3.4  &                     &                      &-209.91&3.20$\sim$3.74&0.24$\sim$72.97&                     &                      &-209.40&3.20$\sim$3.74&0.71$\sim$73.41\\
&&$\rho,~\omega$&$\beta g_V$               &5.2  &                     &                      &150.08&4.68$\sim$5.62&52.77$\sim$0.26&                     &                      &149.69&4.68$\sim$5.62&53.25$\sim$0.72\\
&&$\rho,~\omega$&$\lambda g_V$~(GeV$^{-1}$)&3.133&                     &                      &-113.27&2.82$\sim$3.45&3.82$\sim$50.82&                     &                      &-109.99&2.82$\sim$3.45&4.98$\sim$51.12\\
&&\multicolumn{3}{c@{\vrule width 0.75pt}}{kinetic energy $\langle T\rangle$ (MeV)}&&&236.98&$\cdots$&$\cdots$&&&236.67&$\cdots$&$\cdots$\\
\bottomrule[1pt]
\bottomrule[1pt]
\end{tabular}
\end{table*}

\subsection{Sensitivity of binding energies to the coupling constants}
In addition to the cutoff $\Lambda$, the coupling constants which determine the strength of the potentials are also important to determine whether the three mesons can form bound states. In Eq.~(\ref{potential}), there are five coupling constants, i.e., $g$, $g_\sigma$, $g_V$, $\beta$, and $\lambda$. Among them, only the coupling constant $g$ is determined by the experimental partial decay width $D^*\to D\pi$. All the others are taken from models. For example,  $g_\sigma$ is estimated by the quark model~\cite{Riska:2000gd}, and $\beta$ is obtained from the vector meson dominance mechanism~\cite{Isola:2003fh}. Since there exist uncertainties for the involved coupling constants, it is relevant to study the sensitivity of our results to the adopted values of the coupling constants. We notice that the $\rho$ and $\omega$ exchange potentials share a common coupling constant $g_V$. In order to study the sensitivity of the bound state solutions to the coupling constants, we introduce an about 10\% uncertainty to them, which is somehow arbitrary but nevertheless reasonable.

The numerical results for the $D^*D^*D$ and $D^*D^*D^*$ systems are presented in Tables~\ref{undetermineddsdsd} and~\ref{undetermineddsdsds}, respectively.  We note that the binding energies are highly dependent on the square of the coupling constants, which  is easy to understand since all the potentials in Eq.~(\ref{potential}) are proportional to the square of the coupling constants. Meanwhile, since the changes of the coupling constants could be viewed as perturbations to the potentials, we  estimate the binding energies in perturbation theory. Here, we employ the $S$-wave only $D^*D^*D$ system with $I(J^P)=\frac{1}{2}(0^-)$ as an example to illustrate this point. As shown in Table~\ref{undetermineddsdsd}, we obtain a binding energy $B=3.60$ MeV with a cutoff $\Lambda=1.03$ GeV. The expectation value of the potential from the $\pi$ exchange is $-16.81$ MeV. If we allow $g$ to vary by 10\% (0.9$\sim$1.1), the square of the ratio is in the range 0.81$\sim$1.21. Then the expectation value of the $\pi$ exchange potential is estimated to be in the range of  $-13.63\sim-20.34$ MeV. The resulting binding energy is then  0.41$\sim$7.13 MeV, which is consistent with the exact result 1.08$\sim$7.92 MeV. Following the same approach, when  $g_\sigma$, $\beta g_V$, and $\lambda g_V$ are varied by 10\%, the estimated binding energies in leading order perturbation theory are in the ranges of 0.57$\sim$6.95 MeV, 2.84$\sim$4.44 MeV, 3.29$\sim$3.94 MeV, respectively, which are all consistent  with the exact values 1.27$\sim$7.87 MeV, 2.87$\sim$4.47 MeV, 3.30$\sim$3.95 MeV, respectively.

The message from the above sensitivity study is that although the exact binding energies are sensitive to the values of the coupling constants, the overall picture remains unchanged, i.e., whether there exist some good three-body hadronic molecules.

\section{Summary}\label{summary}

In recent years, the LHCb Collaboration achieved remarkable success in discovering new hadronic states, including many of exotic ones, which cannot fit into the conventional quark model. These observations enriched the members of the exotic hadronic family and improved our understandings of nonperturbative strong interactions. Very recently, the LHCb Collaboration observed a new state $T_{cc}^+$ in the $D^0D^0\pi^+$ channel \cite{LHCb:2021auc}. The $T_{cc}^+$ state could  well be interpreted as a $DD^*$ molecular state and it is the first  double-charm exotic state ever observed.

The observation of the $T_{cc}^+$ state enabled  us  to derive the interaction between two charmed mesons. In Ref.~\cite{Wu:2021kbu}, by reproducing the binding energy of  $T_{cc}^+$, we determined the cutoff $\Lambda$ in the OBE model. This allows us to study hadronic molecular states composed of several charmed mesons. In this work, we studied the existence of  triple-charm molecular states composed of $D^*D^*D$ and $D^*D^*D^*$. Using  the cutoff $\Lambda$ obtained by the binding energy of the $T_{cc}^+$, we find that the $I(J^P)=\frac{1}{2}(0^-,1^-,2^-)$ $D^*D^*D$ and $I(J^P)=\frac{1}{2}(0^-,1^-,2^-,3^-)$ $D^*D^*D$ systems have loosely bound state solutions, which could be viewed as good hadronic molecular candidates. We suggest to search for the $D^*D^*D$ and $D^*D^*D$ molecular states in the following decay modes:
\begin{enumerate}[a.]
\item a double-charm molecular state and a charmed meson,
\item three charmed mesons,
\item three charmed mesons together with a number of pions and photons.
\end{enumerate}
On the other hand, we find that the $I(J^P)=\frac{3}{2}(0^-,1^-,2^-)$ $D^*D^*D$ and $I(J^P)=\frac{3}{2}(1^-,2^-,3^-)$ $D^*D^*D^*$ systems are more difficult to form bound states.

The present framework can be extended to study the $BB^*B^*$-$B^*B^*B^*$ and $BBB^*$ systems. The former has been studied in Ref.~\cite{Garcilazo:2018rwu} and a bound state with $I(J^P)=1/2(2^-)$ and a binding energy of 90 MeV below  the lowest strong decay threshold was found. The latter has been studied in Ref.~\cite{Ma:2018vhp}, where loosely bound states were found for both $I=\frac{1}{2}$ and $I=\frac{3}{2}$. The three-body systems studied in Ref.~\cite{Garcilazo:2018rwu} are  similar to those of  this work, but the number of bound state solutions is far fewer than that obtained in this work. It should be noted that in the present work, we deduced the meson-meson potentials in the one-boson-exchange model, while in Ref.~\cite{Garcilazo:2018rwu}, the two-body interactions are deduced from the $t$ matrices of Refs.~\cite{Vijande:2009kj,Vijande:2009zs,Carames:2012th,Garcilazo:2017ifi}. The different meson-meson potentials are responsible for the different three-body results. In future experiments, searching for  hadronic molecular candidates could help distinguish the different meson-meson interactions.

It is no doubt that the LHCb Collaboration has played an important role in searches for exotic states. The observation of the $T_{cc}^+$ state once again shows the capability of the LHCb detector in this area. With anticipated data accumulation~\cite{LHCb:2018roe},  more exotic states can be expected in the future.

\section*{ACKNOWLEDGEMENTS}

This work is partly supported by the National Natural Science Foundation of China under Grants No.11735003, No.11975041, No.12147152, No.11961141004, and the fundamental Research Funds for the Central Universities. Ming-Zhu Liu acknowledges support from the National Natural Science Foundation of China under Grant No.1210050997. XL is supported by the China National Funds for Distinguished Young Scientists under Grant No. 11825503, National Key Research and Development Program of China under Contract No. 2020YFA0406400, the 111 Project under Grant No. B20063, the National Natural Science Foundation of China under Grant No. 12047501, and the Fundamental Research Funds for the Central Universities.

\end{document}